\documentclass[a4,10pt]{article}
\usepackage{amssymb}
\usepackage{stmaryrd}
\title{Introduction to linear logic and ludics, Part I}
\author{Pierre-Louis Curien (CNRS \& Universit\'e Paris VII)}

\begin{document}

\maketitle

\newtheorem{theorem}{Theorem}[section]
\newtheorem{definition}[theorem]{Definition}
\newtheorem{lemma}[theorem]{Lemma}
\newtheorem{corollary}[theorem]{Corollary}
\newtheorem{proposition}[theorem]{Proposition}
\newtheorem{example}[theorem]{Example}
\newtheorem{exercise}[theorem]{Exercise}
\newtheorem{hardexercise}[theorem]{Exercise${}^*$}
\newtheorem{notation}[theorem]{Notation}
\newtheorem{remark}[theorem]{Remarká}

\newenvironment{branch}{\left\{\begin{array}{l}}{\end{array}\right.}

\newcommand{\qed}{\hfill$\Box$}
\newcommand{\qedm}{\mbox{}\\[-2.5em] \mbox{}\hfill $\Box$} 
\newcommand{\Proof}{\noindent {\sc Proof}. }
\newcommand{\Proofhint}{\noindent {\sc Proof (indication)}. }
\newcommand{\Proofitem}[1]{\medskip \noindent $#1\;$}
\newcommand{\Proofitemf}[1]{ $#1\;$}

\newcommand{\guil}[1]{``#1"}
\newcommand{\lbd}{\lambda}
\newcommand{\mand}{\mbox{ and }}
\newcommand{\mor}{\mbox{ or }}
\newcommand{\nin}{\not\in}
\newcommand{\inc}{\subseteq}
\newcommand{\set}[1]{\{#1\}}
\newcommand{\setc}[2]{\set{#1 \mid #2}}
\newcommand{\faux}{{\bf F}}
\newcommand{\et}{\wedge}
\newcommand{\ou}{\vee}
\newcommand{\vide}{\emptyset}
\newcommand{\implies}{\Rightarrow}
\newcommand{\si}{\Leftarrow}
\renewcommand{\iff}{\Leftrightarrow}
\newcommand{\qqs}[2]{\forall\, #1\;\: #2}
\newcommand{\xst}[2]{\exists\, #1\;\: #2}
\newcommand{\nxst}[2]{\not\!\exists\, #1\;\: #2}
\newcommand \seql[3]{\raisebox{3ex}{$\mbox{#1}\;\;$} \; \shortstack{$#2$ \\ \mbox{}\\
                    \mbox{}\hrulefill\mbox{}\\ \mbox{}\\ $#3$}}
\newcommand \seq[2]{\shortstack{$#1$ \\ \mbox{}\\
                    \mbox{}\hrulefill\mbox{}\\ \mbox{}\\ $#2$}}

\newcommand{\forces}{\makebox[5mm]{\,$\|\!-$}}%

\newcommand{\fun}{\:\rightarrow\:} 
\newcommand{\funt}{\rightarrow^{\ast}} 
\newcommand{\lfun}[1]{\stackrel{#1}{\longrightarrow}}
\newcommand{\lfunt}[1]{\stackrel{#1}{\longrightarrow}\,\!^{\ast}}
\newcommand{\length}[1]{ | #1 |}
\newcommand{\union}{\cup}		
\newcommand{\inter}{\cap}		
\newcommand{\Union}{\bigcup}		
\newcommand{\Inter}{\bigcap}		
\newcommand{\join}{\vee}		
\newcommand{\JOIN}{\bigvee}		
\newcommand{\meet}{\wedge}		
\newcommand{\MEET}{\bigwedge}		

\newcommand{\inv}[1]{#1^{-1}}
\newcommand{\card}{\sharp}		
\newcommand{\pfun}{\rightharpoonup}  
\newcommand{\ul}[1]{\underline{#1}}	

\newcommand{\id}{{\it id}}
\newcommand{\catch}{{\it catch}\;}
\newcommand{\throw}{{\it throw}\;}
\newcommand{\Alt}{ \mid\!\!\mid  } 
\newcommand{\Sub}[3]{#1[#2\leftarrow #3]}	
\newcommand{\sub}[2]{[#2 / #1]}
\newcommand{\subn}[2]{[#2 /_{\!\circ} #1]}
\newcommand{\BO}[4]{#1\,,#2\stackrel{#3}{\Longrightarrow} #4}
\newcommand{\bt}[1]{{\it BT}(#1)}
\newcommand{\lelong}[2]{#1\!\mid_{#2}}	
\newcommand{\dl}{[\![} 			
\newcommand{\dr}{]\!]} 			
\newcommand{\newv}{{\it new}\:}
\newcommand{\deref}{{\it deref}\:}
\newcommand{\nat}{{\bf int}}
\newcommand{\varimp}{{\bf var}\:}
\newcommand{\comm}{{\bf comm}}
\newcommand{\world}{{\bf W}}
\newcommand{\comp}{\circ}
\newcommand{\run}{{\it run}}
\newcommand{\done}{{\it done}}
\newcommand{\readv}{{\it read}}
\newcommand{\writev}[1]{{\it write}(#1)}
\newcommand{\OK}{{\it OK}}
\newcommand{\lpar}{\bindnasrepma}
\newcommand{\with}{\&}
\newcommand{\limpl}{\multimap}
\newcommand \coh {\mathrel{\raisebox{-.3em}{\shortstack{$\frown$\\$\smile$}}}}
\newcommand \icoh {\mathrel{\raisebox{-.3em}{\shortstack{$\smile$\\$\frown$}}}}
\newcommand \seqdbl[2]{\shortstack{$#1$ \\ \mbox{}\\
                    \mbox{}\hrulefill\mbox{}\\  \mbox{}\hrulefill\mbox{}\\ \mbox{}\\ $#2$}}

\newcommand{\bbot}{\bot}
\newcommand{\rst}[1]{\!\upharpoonright^{#1}} 
\newcommand{\cat}[1]{{\bf #1}}
\newcommand{\kleisli}[1]{\kappa(#1)}
\newcommand{\pair}[2]{\langle #1 , #2 \rangle}

\begin{abstract}
In this two-part survey paper, we introduce linear logic and ludics, which were both introduced by Girard, in 1986 and 2001, respectively. They offer a thorough revisitation of mathematical logic from first principles, with inspiration from and applications to computer science.  Some important keywords are: resources, geometry of computation, polarities, interaction, and games.
\end{abstract}

\subsection*{Prerequisites} We assume some basic knowledge of sequent calculus and natural deduction (see \cite{Giraflor, GouMa}), of $\lbd$-calculus (the classical reference is
\cite{Barendregt84}, see also \cite{HindleySeldin86}), and of category theory (only in section \ref{cat-mod}, see \cite{MacLane71}).

\subsection*{Warning} This paper is not trying to survey the whole and impressive body of works on linear logic since 17 years, but rather chooses a route that tries to highlight the deep originality of the subject. The bibliography is in consequence very partial (also reading the latter as a French word, wich meanss something like ``personal'', possibly biased).

\section{Introduction}
Linear logic arose from the semantic study of $\lambda$-calculus, specifically, the second-order typed $\lbd$-calculus, or system F (see \cite{Giraflor}). Shortly after proposing the first denotational model of system F \cite{Gir86}, Girard realized that in coherence spaces (to be introduced in section \ref{cat-mod}) the interpretation of type $A\rightarrow B$ decomposes naturally as
$(!A)\multimap B$ (this will be detailed in section \ref{coh-spaces}), and in almost no time a full-fledged new logic, with a totally innovative way of representing proofs and to execute them, that is, to perform cut-elimination, was created. We tell this story here, not in the order in which it unfolded,
but starting from so-called substructural considerations. We raise the question of what happens when we remove some of the structural rules from sequent calculus, most importantly contraction.
Substructural logics of various kinds have been considered much before the birth of linear logic,
but what characterizes truly linear logic is not that aspect, but rather the controlled way in which the structural rules of weakening and contraction are reintroduced, by means of the new modality 
$!$ (``of course'') (and its dual $?$ (``why not'')). This, together with proof nets, constitutes the major break-through of linear logic.

Let us consider the implication in the absence of contraction. In $\lbd$-calculus, this corresponds to considering terms in which each variable occurs at most once.  (Here we use the Curry-Howard correspondence between proofs and programs, see \cite{Giraflor}).  Consider thus the set of {\em affine} $\lbd$-terms built with the following constraints: $x$ is an affine term, if $M$ is an affine term then $\lbd x.M$ is an affine term, and if $M,N$ are affine terms {\em with disjoint sets of free variables} then $MN$ is an affine term. The key fact about affine terms is the following:
\begin{center}
{\em Affine terms normalize in linear time.}
\end{center}
The reason is trivial: each $\beta$-reduction reduces the size, keeping in mind
that $\Sub{P}{x}{Q}$ is the result of the substitution of at most one occurrence of $x$ with $Q$, so  $\Sub{P}{x}{Q}$ is shorter by at least the two symbols of abstraction and application that have
been absorbed by the reduction of $(\lbd x.P)Q$.  Of course, this linearity is one of the reasons 
for the name Linear Logic. 

In contrast, arbitrary $\lambda$-terms may take an exponential time and more to reach their normal form. For example, the reader may play with $\ul{m}\:\ul{n}$, where $\ul{n}=\lbd fx.f^n(x)$
(Church numerals -- here, $f^0(x)=x$ and $f^{n+1}(x)=f(f^n(x))$). 

As another instance of the phenomenon of duplication, or of {\em multiple use} of a variable (or assumption), think of a constructive reading of $(\qqs{x}{\xst{y}{x<y}})$ from which one may wish to extract a program creating an infinite strictly increasing sequence of numbers. The program extracted from the formula will actually be a procedure to produce a $y>x$, when given an $x$, and it will be called in a loop. In linear logic, this formula would have to be written 
$!(\qqs{x}{\xst{y}{x<y}})$.

\medskip
Rather than continuing a long introduction, we decide to enter in  the subject proper, and we content ourselves with giving a plan of the sections to come. In section \ref{mall}, we introduce so-called {\em multiplicative-addditive} linear logic, and we complete the description of (propositional) linear logic with the 
{\em exponentials} in section \ref{full-ll}. We shall not consider quantifiers in this survey paper, in any case, they look very much as usual. Then, in section \ref{undecidable}, we illustrate the expressivity of linear logic by 
sketching the proof that propositional linear logic is undecidable, a property that sharply distinguishes linear logic from classical and intuitionistic logics. The following three sections
are concerned with the semantics of linear logic. We first discuss phase semantics, which is just a semantics of provability  (section \ref{phase-sem}), then we introduce the coherence semantics, already alluded to (section \ref{coh-spaces}), and finally we address the question of more general categorical definitions of models (section \ref{cat-mod}). 

In part II, we shall go back to syntactic issues and introduce proof nets. In particular, we shall
address the so-called correctness criteria: proof nets are elegant representations of proofs avoiding some useless sequentializations imposed by sequent calculus. Reconstructing the
sequentializations and thus asserting that a proof structure, that is, a would-be proof net, is actually a proof net, is a fascinating problem, that relates to games semantics
\cite{Hyland97, AbraMcCus99}. 
The last sections (a polished form of \cite{Cu01}) are devoted to ludics, which while strongly inspired by linear logic
takes a more radical view of proofs as interactive processes, evolving in {\em space} and time.

\medskip
Here is a table of sources I have been using for writing this paper. The reader will be able to find more complete information and proofs by going to these references, as well as to the many others that he will find in their bibliographies or on the Web!
\begin{center}
\begin{tabular}{lll}
sections \ref{mall} and  \ref{full-ll} & &  \cite{Gir87,GirLLSS,DD} (slightly revisited here in the light of ludics)\\
section \ref{undecidable}& & \cite{LMSDec} (a remarkably well-written paper)\\
sections \ref{phase-sem} and \ref{coh-spaces}&&  \cite{Gir87} \\
section \ref{cat-mod}&& \cite{Mellies03} (which itself is a survey)\\
proof nets&& \cite{Gir87,DR89, GueMa2000,Baillot99}\\
ludics && \cite{LS00}
\end{tabular}
\end{center}

\section{Multiplicatives and additives} \label{mall}
In sequent calculus, depending on presentations, you can see the left and right rules for, say, conjunction given in this format:
$$\begin{array}{llll}
\seq{\Gamma,A\vdash \Delta}{\Gamma,A\meet B\vdash \Delta} \quad \seq{\Gamma,B\vdash \Delta}{\Gamma,A\meet B\vdash \Delta} &&&
\seq{\Gamma\vdash A,\Delta\quad\Gamma\vdash B,\Delta}{\Gamma\vdash A\meet B,\Delta}
\end{array}$$
or in that format:
$$\begin{array}{llll}
\seq{\Gamma,A,B\vdash \Delta}{\Gamma,A\meet B\vdash \Delta}  &&&
\seq{\Gamma_1\vdash A,\Delta_1\quad\Gamma_2\vdash B,\Delta_2}{\Gamma_1,\Gamma_2\vdash A\meet B,\Delta_1,\Delta_2} 
\end{array}$$
The two formats have been called {\em additive} and {\em multiplicative}, respectively, by Girard.
We recall that a sequent $\Gamma\vdash\Delta$ is made of two multisets of formulas (that is, we work modulo the so-called exchange rule which says that $\Gamma\vdash\Delta$ entails any of its permutations -- on the contrary, rejecting this rule leads to a complicated and still not very well-understood subject called non-commutative (linear) logic). The notation $\Gamma_1,\Gamma_2$ stands for the multiset union of $\Gamma_1$ and $\Gamma_2$.  

Logicians did not bother about these different presentations, because they are equivalent. Yes, but modulo the contraction and weakening rules:
$$\begin{array}{llllll}
\seq{\Gamma,A,A\vdash \Delta}{\Gamma,A\vdash\Delta} &
\seq{\Gamma\vdash A,A,\Delta}{\Gamma\vdash A,\Delta} &&&
\seq{\Gamma\vdash\Delta}{\Gamma,A\vdash\Delta} & \seq{\Gamma\vdash\Delta}{\Gamma\vdash A,\Delta}
\end{array}$$
Let us show this, say, for the left introduction rule. Let us first derive the multiplicative format
from the additive one:
$$\seq{\seqdbl{\Gamma_1\vdash A,\Delta_1}{\Gamma_1,\Gamma_2\vdash A,\Delta_1,\Delta_2}\quad\seqdbl{\Gamma_2\vdash B,\Delta_2}{\Gamma_1,\Gamma_2\vdash B,\Delta_1,\Delta_2}}{\Gamma_1,\Gamma_2\vdash A\meet B,\Delta_1,\Delta_2}$$
where the double line expresses multiple use of the weakening rules. For the converse direction we have:
$$\seq{\Gamma\vdash A,\Delta\quad\Gamma\vdash B,\Delta}{\seqdbl{\Gamma,\Gamma\vdash A\meet B,\Delta,\Delta}{\Gamma\vdash A\meet B,\Delta}}
$$
using contractions. If we remove contraction and weakening, then the presentations are no longer equivalent and hence define {\em different} connectives, $\otimes$ and $\&$, respectively called ``tensor'' and ``with''.
Indeed, you will see in section \ref{coh-spaces} that they are interpreted very differently. Moreover, the coherence space model (to which the curious reader may jump directly)
enlightens the terminology  multiplicative--additive, as the reader will check that the cardinal
of (the interpretation of) $A\otimes B$ (resp. $A\& B$) is the {\em product} (resp. the {\em sum}) of the cardinals of (the interpretations of) $A,B$.

Similarly, we split ``disjunction'' in two connectives, $\lpar$ and $\oplus$, respectively called ``par'' and ``plus''.  As usual in sequent calculus, the left rule for, say, $\otimes$, tells us by duality what the right rule for
$\lpar$ is, and similarly for $\oplus$. This leads us to the following monolateral sequent presentation of the four connectives:

$$\begin{array}{llclc}
\mbox{\bf MULTIPLICATIVES}\\
 && \seq{\vdash A,B, \Gamma}{\vdash A\lpar B, \Gamma}  &&
\seq{\vdash A,\Gamma_1\quad\vdash B,\Gamma_2}{\vdash A\otimes B,\Gamma_1,\Gamma_2} 
 \\\\
\mbox{\bf ADDITIVES}\\
&& \seq{\vdash A, \Gamma}{\vdash A\oplus B, \Gamma} \quad \seq{\vdash B,\Gamma}{\vdash A\oplus B,\Gamma} &&
\seq{\vdash A,\Gamma\quad\vdash B,\Gamma}{\vdash A\& B,\Gamma}
\end{array}$$

Considered as (associative and commutative) operations, these four connectives 
have units ``true'' and ``false'', each of them splits into the multiplicative and additive one:

$$\begin{array}{lllll}
 \lpar & \otimes && \oplus & \with\\
 \bot & 1 && 0 & \top
\end{array}$$
(As for the terminology and for mnemotechnicity, the names  1 and  0 correspond to the usual multiplicative and additive notation in groups.)  
We shall synthesize the rules for the units from the requirement that the isomorphisms $A\otimes 1\approx A$, etc..., should hold. In order to prove $\vdash A\otimes 1,\Gamma$ from $\vdash A,\Gamma$ we should have $\vdash 1$. In order to prove $\vdash A\lpar \bot,\Gamma$ from $\vdash A,\Gamma$, we should have $\vdash A,\bot,\Gamma$ as an intermediate step, which suggests the rule for $\bot$ given below.  We derive $\vdash A\oplus 0,\Gamma$ from $\vdash A,\Gamma$ by the  (left)  $\oplus$ rule, without any need for a (right) rule for $0$. Finally, we need $\vdash \top,\Gamma$  to derive $\vdash A\with\top,\Gamma$ from $\vdash A,\Gamma$:

$$\begin{array}{lllllll}
\seq{\vdash A\quad\vdash 1}{\vdash A\otimes 1} && \seq{\seq{\vdash A,\Gamma}{\vdash A,\bot,\Gamma}}{\vdash A\lpar\bot,\Gamma} && \seq{\vdash A,\Gamma}{\vdash A\oplus 0,\Gamma} && \seq{\vdash A,\Gamma\quad\vdash \top,\Gamma}{\vdash A\&\top,\Gamma}
\end{array}$$
The rules are thus:

$$\begin{array}{llllllllllll}
\mbox{\bf UNITS}\\
&& \seq{}{\vdash 1} &&& \seq{\vdash\Gamma}{\vdash\bot,\Gamma} &&& \mbox{no rule for }0 &&& \seq{}{\vdash\top,\Gamma}
\end{array}$$

\smallskip\noindent
An even shorter way to synthesize the rules for the units is to formulate rules for $n$-ary versions of the four connectives, and then to take the degenerate case $n=0$. The assumption of the $n$-ary $\lpar$ rule is $\vdash A_1,\ldots,A_n,\Gamma$, which degenerates to $\vdash\Gamma$. For the tensor rule there are $n$ assumptions and the conclusion is
$\vdash A_1\otimes\ldots\otimes A_n,\Gamma$, with $\Gamma=\Gamma_1,\ldots,\Gamma_n$, and in the degenerate case we thus have no assumption and $\Gamma=\emptyset$.  There are $n$ $\oplus$ rules, and thus no rule in the degenerate case. Finally, the $\with$ rule has $n$ assumptions, and thus no assumption in the degenerate case.

\medskip
What about negation? In classical logic, negation can be boiled down to negation on atomic formulas only, thanks to De Morgan laws: $\neg (A\meet B)=(\neg A)\join(\neg B)$, from which the other De Morgan law follows, using $\neg\neg A=A$. This works also for linear logic, where negation
is written $A^\bot$ (``$A$ orthogonal'', or ``$A$ perp'').  But the logical operations being different, negation in linear logic is very different from that of classical logic. In classical logic, we have both
$A=\neg\neg A$ and $\vdash A\join\neg A$ (the latter being what constructivists criticized), while in linear logic we have $A=A^{\bot\bot}$ and $\vdash A\lpar A^\bot$, but {\em we do not have}
$\vdash A\oplus A^\bot$, and as a matter of fact linear logic retains the distinctive features
that make intuitionistic logic so nice:  confluence (of cut-elimination), and the disjunction property
(for $\oplus$, see below).  

Thanks to De Morgan laws, we can dispense with negation as a connective, and consider $A^\bot$ as an abbreviation for its De Morgan normal form obtained by applying the following rules (where we have added the not-yet introduced  ``!'' and ``?'' for completeness):
$$\begin{array}{llll}
(A\otimes B)^\bot \rightarrow (A^\bot)\lpar(B^\bot) &
(A\lpar B)^\bot \rightarrow (A^\bot)\otimes(B^\bot) &
1^\bot \rightarrow \bot & \bot^\bot\rightarrow 1\\
(A\with B)^\bot \rightarrow (A^\bot)\oplus(B^\bot) &
(A\oplus B)^\bot \rightarrow (A^\bot)\with(B^\bot) &
\top^\bot \rightarrow 0 & 0^\bot\rightarrow \top\\
(!A)^\bot\rightarrow ?(A^\bot) & (?A)^\bot\rightarrow !(A^\bot)
\end{array}
$$
(Notice that the dual under $^\bot$ of a multiplicative (resp. additive) connective is a multiplicative (resp. additive) connective.) In the sequel, we shall freely use bilateral sequents in examples, but they can be seen as syntactic sugar for monolateral ones, for example $A_1,A_2\vdash B$ 
is a version of $\vdash A_1^\bot,A_2^\bot,B$. Hence, thanks to duality, linear logic puts all formulas of a sequent on the same ground, in  contrast to intuitionistic logic, where there is a clear distinction between input ($\Gamma$) and output ($A$) in a sequent $\Gamma\vdash A$.

\medskip
To complete the description of the multiplicative-additive fragment of linear logic, we need axioms, and the cut rule. The latter is not needed in the sense that it can be eliminated (just as in classical or intuitionisitc sequent calculus), but one can hardly use the system without cuts, which carry
the intelligent part of proofs.
$$\begin{array}{lllllll}
\mbox{\bf AXIOM} &&&& \mbox{\bf CUT}\\
& \seq{}{\vdash A,A^\bot} &&&& \seq{\vdash A, \Gamma_1 \quad\vdash A^\bot,\Gamma_2}{\vdash \Gamma_1,\Gamma_2}
\end{array}$$
(Note that the cut rule is given in a multiplicative style, i.e., it is quite similar to the tensor rule. That the axiom has this most economical format, as opposed to $\vdash A,A^\bot, \Gamma$, should come as no surprise, as the latter embodies an implicit weakening.)

\medskip
As explained in the introduction, the main motivation for the removal of (weakening and) contraction is the control on normalisation. Indeed, the reader can easily check that all affine $\lbd$-terms can be typed by the following rules:
$$\begin{array}{lll}
\seq{}{x:A\vdash x:A} & \seq{\Gamma_1\vdash M_1:A\multimap B\quad\Gamma_2\vdash M_2:A}{\Gamma_1,\Gamma_2\vdash M_1M_2:B} & \seq{\Gamma,x:A\vdash M:B}{\Gamma\vdash \lbd x.M:A\multimap B}
\end{array}$$

\noindent
Notice the multiplicative style of the crucial typing rule for application. As usual, ``$A$ implies $B$'' is ``not $A$ or $B$'', and specifically here $A\multimap B=(A^\bot)\lpar B$ ($\multimap$ reads ``linearly implies'').
 This natural deduction style rule is indeed encoded in sequent calculus using the (multiplicative) cut rule:
$$\seq{\Gamma_1\vdash A^\bot\lpar B\quad\seq{\Gamma_2\vdash A\quad\seq{}{\vdash B^\bot,B}}{\Gamma_2\vdash A\otimes B^\bot,B}}{\Gamma_1,\Gamma_2\vdash B}$$

\noindent
This leads us to the main conceptual novelty of linear logic: the understanding of formulas as {\em resources}, that are {\em consumed} through entailment: in the sequent  $A\vdash B$, the resource
$A$ is consumed in order to {\em get} $B$ (think also of a chemical reaction). This is like in an abstract state machine, which changes state upon performing certain actions. As a matter of fact, we present an encoding of such a machine in section \ref{undecidable}. With this reading in mind, one easily understands how bad contraction and weakening are. Suppose that $B$ is some good, and $A$ is a given amount of money, say 20 Euros. Then $A,A\vdash B$ reads as:  ``you can acquire $B$  against 40 Euros''. The contraction rule would say
that you could acquire the same good for 20 Euros. Thus, contraction is not very good for the merchant.  It is not good in chemistry either: in a reaction, in order to obtain $2H_2O$ you need to have exactly four $H^+$ and two $O^{-2}$.
Another way to criticize the rule is 
by looking  from conclusion to antecedents (proof-search): if you only have $A$ to get $B$, your resource $A$ cannot be magically duplicated, as the contraction rule would imply.

As for weakening, if you can buy $B$ with $A$, as formalized by $A\vdash B$, why would you -- the buyer -- spend an additional amount of money $C$ to get $B$ ($C,A\vdash B$)? Weakening is not good for the client.

\medskip
So far, we have insisted on a dichotomy between additive and multiplicative connectives.
There is however another grouping of connectives, that was sitting there from the beginning, and that has even  been somehow made explicitly by the choice of symbols: $\otimes$ and $\oplus$ on one hand, and $\&$ and $\lpar$ on the other hand. Girard noticed that $\otimes$ distributes over $\oplus$ (like in arithmetic!), and that dually $\lpar$ distributes over $\&$, while other thinkable combinations do not distribute upon each other (it is an instructive exercise to check this). Let us examine the proof of this distributivity law.
We have to prove $\vdash (A\otimes(B\oplus C))^\bot,(A\otimes B)\oplus(A\otimes C)$ and
$\vdash ((A\otimes B)\oplus(A\otimes C))^\bot,A\otimes(B\oplus C)$. Both proofs begin by decomposing the $\lpar$'s and the $\&$'s. For the first sequent, this goes as follows:
$$\seq{\seq{\vdash A^\bot, B^\bot,(A\otimes B)\oplus(A\otimes C)\quad\vdash A^\bot, C^\bot,(A\otimes B)\oplus(A\otimes C)}{\vdash A^\bot, (B^\bot\& C^\bot),(A\otimes B)\oplus(A\otimes C)}}
{\vdash A^\bot\lpar (B^\bot\& C^\bot),(A\otimes B)\oplus(A\otimes C)}$$

\noindent 
and we are left to prove the subgoals $\vdash A^\bot, B^\bot,(A\otimes B)\oplus(A\otimes C)$ and
$\vdash A^\bot, C^\bot,(A\otimes B)\oplus(A\otimes C)$.
Similarly, the search of a  proof of the second sequent yields
$\vdash A^\bot, B^\bot,A\otimes(B\oplus C)$ and $\vdash A^\bot, C^\bot,A\otimes(B\oplus C)$ as subgoals. Before we proceed, note an important feature of $\with$ and $\lpar$: in each of the two proof constructions, the two subgoals together are actually equivalent to the original goal: these two connectives are {\em reversible} -- an affirmation that we shall make formal just below.
In contrast, in order to complete the proof of any of the subgoals, we have to make (irreversible) decisions. We just give the proof of  $\vdash A^\bot, B^\bot,(A\otimes B)\oplus(A\otimes C)$, but the others are similar:
$$\seq{\seq{\seq{}{\vdash A^\bot,A}\quad\quad\seq{}{\vdash B^\bot,B}}{\vdash A^\bot, B^\bot,(A\otimes B)}}{\vdash A^\bot, B^\bot,(A\otimes B)\oplus(A\otimes C)}$$

\noindent
The two decisions which we have made in this proof are: to choose the left $\oplus$ rule, and
then to split $A^\bot,B^\bot$ in the way we did, assigning the resource $A^\bot$ for the proof of
$A$ and $B^\bot$ for the proof of $B$.   

The opposition $\otimes$--$\oplus$ versus $\lpar$--$\&$ is actually fundamental, and has played an important role in the genesis of ludics (see part II). The following table summarizes its significance:
\begin{center}
\begin{tabular}{ccc}
$\otimes\quad\oplus$ && $\lpar\quad\&$\\\\
Irreversible && Reversible\\
Non-deterministic && Deterministic\\
Active && Passive\\
Player && Opponent\\
External choice && Internal choice\\
Positive && Negative
\end{tabular}
\end{center}
We first show that the rules for $\lpar$ and $\&$ are reversible, that is, that the conclusions of the rules hold if {\em and only if} their antecedents hold, as follows:
$$\begin{array}{llll}
\seq{\vdash\Gamma,A\lpar B\quad\seq{\seq{}{\vdash A^\bot,A}\quad\seq{}{\vdash B^\bot,B}}{\vdash A^\bot\lpar B^\bot,A,B}}{\vdash\Gamma,A,B} &&&
\seq{\vdash A\&B,\Gamma\quad\seq{\seq{}{\vdash A^\bot,A}}{\vdash A^\bot\oplus B^\bot,A}}{\vdash A,\Gamma}
\end{array}$$

\noindent Hence we lose nothing by replacing $\vdash\Gamma,A\lpar B$ with $\vdash\Gamma,A,B$ in proof-search, and moreover we can gain something, since
the liberated immediate subformulas $A$ and $B$ can then be sent to different antecedents when decomposing a tensor (as in the proof of $\vdash A\lpar B,A^\bot\otimes B^\bot$).
The $\lpar$ and $\&$ rules are moreover {\em deterministic} in the sense that, once the formula to be decomposed in the sequent is fixed, then there is no choice on how to do it: for the $\lpar$ rule, just dissolve the $\lpar$, and for the $\&$ rule, prove the two subgoals, which are the same sequent up to the replacement of the formula by one of its immediate subformulas.
As we shall see in part II, provided a certain proof-search discipline is respected, which 
can be phrased as ``apply reversible rules in priority'', and if maximal groups of rules 
concerning reversible (resp. irreversible) connectives are grouped and considered as a single {\em synthetic rule},
then sequents always contain at most
one formula whose topmost connective is  reversible, and therefore even the choice of which formula of the sequent to decompose is deterministic. This absence of initiative can be termed {\em passivity}.  

 Let us examine a contrario what makes the other connectives {\em active} and {\em irreversible}. We have already hinted at this when we proved the distributivity law. Each $\oplus$ rule, read bottom-up,
chooses one of $A$ or $B$, while each instance of the $\otimes$-rule makes a choice of how to split the (rest of the) sequent  in two parts. The choice of which formula of the sequent to decompose is also non-deterministic. In other words, the two connectives are associated with some actions that must be taken.

\medskip
The next pair in our list places us in the tradition of the {\em dialogue game} interpretation of proofs, which goes back to Gentzen. A proof is the work of one person, the {\em Player} (think of a student), which another
person, the {\em Opponent} (think of an examiner)  can check.  Checking goes by tests, or {\em plays}.
The Opponent challenges the conclusion of the proof, to which the Player has to answer by displaying the last rule he used. The Opponent then chooses one of the antecedents of the rule, and challenges the Player to justify that particular antecedent, to which the Player answers by 
displaying the last rule used to show this antecedent, etc... The play results in disclosing, or exploring, a part of the proof. We shall come back to this game interpretation in part II. Here, we shall content ourselves with an illustration, Yves Lafont's menu:
\begin{center}
\begin{tabular}{ccc}
\begin{tabular}{c}
{\bf Menu (price 17 Euros)}\\\\
{\it Quiche or Salad}\\
{\it Chicken or Fish}\\
{\it Banana or ``Surprise du Chef$^*$}''\\\\
{\small (*) either ``{\it Profiteroles}'' or ``{\it Tarte Tatin}''}
\end{tabular} && $17E\vdash\left\{\begin{array}{c}
(Q\& S)\\\otimes\\
(C\&F)\\\otimes\\
(B\&(P\oplus T))
\end{array}\right.$
\end{tabular}
\end{center}
We can recognize here some of the meanings that we already discussed. The right of the sequent is what you can get for $17$ Euros. The tensor tells that for this price you get one ``entr\'ee'', one dish and one dessert. The difference between $\&$ and $\oplus$ is a bit more subtle, and the 
game interpretation is helpful here. So let us start again from the beginning, considering a play between the restaurant manager (the Player) and the customer (the Opponent). It is the Player's responsibility to split the 17$E$ into three parts, corresponding to the cost of the three parts of the
meal. May be, this is done as follows:
$$\seq{5E\vdash
Q\& S\quad 8E\vdash C\&F\quad
4E\vdash B\&(P\oplus T)}{17E\vdash
(Q\& S)\otimes(C\&F)\otimes
(B\&(P\oplus T))}$$
Now let the Opponent challenge $5E\vdash Q\& S$:
$$\seq{5E\vdash Q\quad 5E\vdash S}{5E\vdash Q\&S}$$
which reads as: both Quiche and Salad are available to the customer, but he can get only one, and it is {\em his} choice of picking one of the antecedents and to order, say,  a Quiche. Thus the additive conjunction can be understood as a ... disjunction embodying a notion of {\em external choice}  (remember that in our example the customer is the Opponent, or the context, or the environment). Let us now analyse a proof of $4E\vdash B\&(P\oplus T)$:
$$\seq{4E\vdash B\quad \seq{4E\vdash T}{4E\vdash P\oplus T}}{4E\vdash B\&(P\oplus T)}$$
Suppose that the Opponent chooses the Surprise. Then it is the Player's turn, who justifies
$4E\vdash P\oplus T$ using the right $\oplus$ rule. So, the Opponent will get a Tarte Tatin, but 
the choice was in the Player's hands. Thus $\oplus$ has an associated meaning of {\em internal choice}. In summary, two forms of choice, external  and internal, are modelled by $\with$ and $\oplus$, respectively.   In the case of $\oplus$, whether $A$ or $B$ is chosen is controlled by the rule, that is, by the Player. In the case of $\&$, the choice between $A$ or $B$ is a choice of one of the antecedents of the rule, and is in the hands of the Opponent.

Actually, our image becomes even more acurate if we replace the customer with an inspector  (in summer, many restaurants propose unreasonable prices to the tourists...). The inspector will not consume the whole menu, he will just check (his choice!) whether what is offered, say for the entr\'ee, is correct (not over-priced, fresh enough...).
Another inspector, or the same inspector, may want to do another experiment later, checking this time on dessert: using this sharper personification, the game as explained above is more fully reflected.

\medskip All these oppositions confirm a fundamental {\em polarity}: by convention, we shall term $\&$ and $\lpar$ as {\em negative}, and $\otimes$ and $\oplus$ as {\em positive}.

\section{ Full linear logic} \label{full-ll}

In order to recover the power of the full $\lambda$-calculus (or of intuitionistic logic), we reintroduce the structural rules, but only on formulas marked with a modality (actually two: ``!'' and its dual ``?''):

$$\begin{array}{lll}
\seq{\vdash ?A,?A,\Gamma}{\vdash ?A,\Gamma} && \seq{\vdash\Gamma}{\vdash ?A,\Gamma}
\end{array}$$

\noindent
The rules are easier to explain in a bilateral format. If you want to prove $!A,\Gamma\vdash B$, then contraction allows you to give yourself the freedom of using the assumption $!A$ twice, or more: if you have been able to prove $!A,\ldots,!A,\Gamma\vdash B$, then you can conclude by using (repeated contractions). Similarly, weakening allows you to forget assumptions of the form $!A$:  having a proof of $\Gamma\vdash B$ is enough to get  $!A,\Gamma\vdash B$.

The connective $!$ is a connective just as the others, and thus has its left and right introduction rules ($?\Gamma$ denotes a multiset of formulas each of the form $?A$):

$$\begin{array}{llllllll}
{\bf Dereliction} &&&&& {\bf Promotion}\\
& \seq{\Gamma,A}{\vdash\Gamma,?A} &&&&& \seq{\vdash ?\Gamma,A}{\vdash ?\Gamma,!A}
\end{array}$$

\noindent
Our description of linear logic (LL) is now complete! The two connectives $!$ and $?$ are called the {\bf EXPONENTIALS}.

\medskip
A variation on the promotion rule, which is inspired by categorical considerations (see section \ref{cat-mod}), and also by considerations of the {\em geometry of interaction} (see part II) is:

$$\begin{array}{lllll}
\seq{\vdash \Gamma,A}{\vdash ?\Gamma,!A} &&&& \seq{\vdash ??A,\Gamma}{\vdash ?A,\Gamma}
\end{array}$$

\noindent
These two rules together are equivalent to the promotion rule. We just show here how  the second rule can be derived:
$$\seq{\vdash ??A,\Gamma\quad\seq{\seq{}{\vdash !A^\bot,?A}}{\vdash !!A^\bot,?A}}{\vdash ?A,\Gamma}$$

\noindent For intuitionistic or classical sequent calculus, the key result to establish in order to
 be able to make some use of it is the cut-elimination theorem. It holds here too, but
we shall not prove it here (this is done in full detail, say, in the appendix of \cite{LMSDec}), and we shall rather content ourselves with giving only a few cut-elimination rules, without conviction, since a much better representation of proofs -- prof nets -- will be given in part II. It is good that the reader see  what he gains, though!
$$\seq{\seq{}{\vdash A^\bot,A}\quad \stackrel{\small\begin{array}{c}\Pi\\\vdots\end{array}}{\vdash A^\bot,\Gamma}}{\vdash A^\bot,\Gamma} \quad\quad\longrightarrow \quad\quad\stackrel{\small\begin{array}{c}\Pi\\\vdots\end{array}}{\vdash A^\bot,\Gamma}$$

$$\seq{\seq{\stackrel{\small\begin{array}{c}\Pi\\\vdots\end{array}}{\vdash\Gamma_1,?B^\bot,?B^\bot}}{\vdash\Gamma_1,?B^\bot}\quad\seq{\stackrel{\small\begin{array}{c}\Pi'\\\vdots\end{array}}{\vdash?\Gamma_2,B}}{\vdash ?\Gamma_2,!B}}
{\vdash \Gamma_1,?\Gamma_2}\quad\quad\longrightarrow\quad\quad
\seqdbl{\seq{\seq{\stackrel{\small\begin{array}{c}\Pi\\\vdots\end{array}}{\vdash\Gamma_1,?B^\bot,?B^\bot}\quad\seq{\stackrel{\small\begin{array}{c}\Pi'\\\vdots\end{array}}{\vdash?\Gamma_2,B}}{\vdash ?\Gamma_2,!B}}
{\vdash\Gamma_1,?\Gamma_2,?B^\bot}\quad \seq{\stackrel{\small\begin{array}{c}\Pi'\\\vdots\end{array}}{\vdash?\Gamma_2,B}}{\vdash ?\Gamma_2,!B}}
{\vdash \Gamma_1,?\Gamma_2,?\Gamma_2}}
{\vdash \Gamma_1,?\Gamma_2}$$

\noindent
The second rule illustrates duplication of proofs, while the first one is the reinsuring one: cuts may vanish!

***** AJOUTER DES REGLES COMMUTATIVES *****

\medskip
We can now give a translation of (simply-typed) $\lambda$-calculus:
$$\begin{array}{lllll}
\seq{}{\Gamma,x:A\vdash x:A} && \seq{\Gamma,x:A\vdash M:B}{\Gamma\vdash \lbd x.M:A\rightarrow B} && \seq{\Gamma\vdash M:A\rightarrow B\quad \Gamma\vdash N:A}{\Gamma\vdash MN:B}
\end{array}$$
 into LL proofs. This translation takes (a proof of) a judgement $\Gamma\vdash M:A$ and turns it into 
a proof $\dl\Gamma\vdash M:A\dr$
of $\vdash ?(\Gamma^\ast)^\bot,A$, where $^\ast$ applies to all formulas of $\Gamma$, (hereditarily) turning all $B\rightarrow C$'s into $?B^\bot\lpar C$. Formally:
$$\begin{array}{lllll}
A^\ast=A\;\;(A\mbox{ atomic}) && (B\rightarrow C)^\ast=?(B^\ast)^\bot\lpar C^\ast && 
?(\Gamma^\ast)^\bot=\setc{?(A\ast)^\bot}{A\in\Gamma}
\end{array}$$

\noindent
The reason why $\Gamma\vdash A$ becomes $\vdash ?\Gamma^\bot,A$ (or $!\Gamma\vdash A$) should be clear: a variable can have several occurrences in a term, whence the need of a contraction rule to type it (using multiplicative and exponential connectives only).
The translation is as follows (we omit the $^\ast$ for a better readability).

$$\begin{array}{llll}
\mbox{\bf Variable} &&& \mbox{\bf Abstraction}\\
\dl\Gamma,x:A\vdash x:A\dr \;=\; \seq{\seqdbl{\seq{}{A^\bot,A}}{\vdash ?\Gamma^\bot,A^\bot,A}}{\vdash ?\Gamma^\bot,?A^\bot,A}&&&
\dl\Gamma\vdash \lbd x.M:A\rightarrow B\dr \;=\;
\seq{\stackrel{\small\begin{array}{c}
\dl\Gamma,x:A\vdash M:B\dr\\
\vdots
\end{array}}{\vdash ?\Gamma^\bot,?A^\bot,B}}{\vdash ?\Gamma^\bot,(?A^\bot\lpar B)}
\end{array}$$

\medskip
$$\begin{array}{l}
\mbox{\bf Application}\\
\dl\Gamma\vdash MN:B\dr \;=\; \seqdbl{
\seq{
\stackrel{\small\begin{array}{c}\dl\Gamma\vdash M:A\rightarrow B\dr\\
\vdots
\end{array}}{\vdash ?\Gamma^\bot,?A^\bot\lpar B}\quad\quad\seq{
\seq{
\stackrel{\small\begin{array}{c}
\dl\Gamma\vdash N:A\dr\\\vdots
\end{array}}{\vdash ?\Gamma^\bot,A}}{\vdash ?\Gamma^\bot,!A}\quad\seq{}{\vdash B^\bot,B}}
{\vdash ?\Gamma^\bot,!A\otimes B^\bot,B}}
{\vdash ?\Gamma^\bot,?\Gamma^\bot,B}}
{\vdash ?\Gamma^\bot,B}
\end{array}$$

\smallskip
\noindent
Of course, this is not enough to declare that we have a satisfactory translation. The translation should be also computationally sound, in the sense that if $M$ reduces to $N$, then the translation of $M$ reduces to the translation of $N$. It is the case with this translation, but it is not
so immediate to establish, as the translation does two things at the same time: it  factors a translation from intuitionistic natural deduction (NJ)  to intuitionistic sequent calculus (LJ) and a translation of LJ to LL. 

There are also other qualities that one may require of a translation: for example, that it should send cut-free proofs  to cut-free proofs.
Another (stronger) requirement is that the translation 
should respect the skeleton of the translated proof, that is, the translation contents itself with introducing some derelictions and promotions in the original proof. We refrain from going into this here. Let us just say that the (LJ to LL part of the) above translation does not respect these two requirements, but other ``more verbose'' ones (i.e., using more occurrences of the modalities) do.
A thorough analysis of the translations of intuitionistic (and also classical) logic in linear logic can be found in
\cite{DJS93,DJS97} (see also \cite{DD}).

\medskip
We end the section with an important isomorphism: $!(A\&B)\equiv (!A)\otimes(!B)$  (from which the name exponential comes: think of ``$e^{A+B}=e^A\times e^B$''). We prove only one side and leave the other to the reader (of course, we are doing the same job as at the beginning  of section \ref{mall}):

$$\seq{\seq{\seq{\seq{\seq{\seq{}{\vdash A^\bot,A}}{\vdash ?A^\bot,A}}{\vdash ?A^\bot,?B^\bot,A}\quad\seq{\seq{\seq{}{\vdash B^\bot,B}}{\vdash ?B^\bot,B}}{\vdash ?A^\bot,?B^\bot,B}}
{\vdash ?A^\bot,?B^\bot,A\& B}}
{\vdash ?A^\bot,?B^\bot,!(A\& B)}}{\vdash ?A^\bot\lpar?B^\bot,!(A\& B)}$$

\section{Proposition linear logic is undecidable} \label{undecidable}

We are interested in the decidability of the following problem: given $\Gamma$, is it decidable wether $\vdash\Gamma$ is provable? For MALL, this problem is decidable,
as the possible cut-free proofs of a given sequent can be enumerated. This is because in a cut-free proof the formulas that can appear in the proof are subformulas of the formulas in the conclusion (this is the well-known {\em subformula property}), so there is a finite number of them, and moreover the absence of contraction guarantees that there are also finitely many possible sequents. 
Think of a proof of $\vdash\Gamma,A$ that would repeatedly apply contraction to $A$, thus searching
$\vdash\Gamma,A,A$, $\vdash\Gamma,A,A,A$,\ldots. This is not allowed in MALL (since the exponentials are omitted). 
But the complexity is high, and in fact the problem is NP-complete. We refer to  \cite{LMSDec}[Section  2] for the instructive and pleasing proof, which relies on an
encoding of quantified boolean formulas in MALL. 

What about full LL? It turns out that 
the Halting Problem for (a variant of) Two Counter Machines can be faithfully encoded in propositional linear logic, which entails the undecidability of the logic \cite{LMSDec}[Section 3]. (Note that this is in sharp contrast with classical propositional logic, for which decidability follows from the possibility of an exhaustive check of the truth values.)
We next explain the encoding, and the structure of the proof of its faithfulness.

A Two Counter Machine (TCM) is given by a set $S$ of {\em states}, with two distinguished states $q_I$ and $q_F$ (the {\em initial} and {\em final} states, respectively) and by 5 sets of {\em instructions}
$I_1,I_2,I_3,I_4,I_5$, where $I_1$ through $I_4$ are subsets of $S\times S$ and $I_5$ is a subset of $S\times S\times S$. We use the following user-friendlier notation:
$$\left.\begin{array}{r}
q_i\lfun{+_A}q_j\\
q_i\lfun{-_A}q_j\\
q_i\lfun{+_B}q_j\\
q_i\lfun{-_B}q_j\\
q_i\lfun{\it fork}q_j,q_k\\
\end{array}\right\}\mbox{ for }
\left\{\begin{array}{l}
(q_i,q_j)\in I_1\\
(q_i,q_j)\in I_2\\
(q_i,q_j)\in I_3\\
(q_i,q_j)\in I_4\\
(q_i,q_j,q_k)\in I_5\\
\end{array}\right.$$
These instructions act on {\em instantaneous descriptions} (ID's), which are multisets $s$ of 
triplets of the form $(q,m,n)\in S\times\omega\times\omega$ ($\omega$ being the set of natural numbers), as follows (where union is taken in the sense of multisets, e.g. $\set{x,y}\union\set{x}=\set{x,x,y}$):
$$\begin{array}{ll}
\seq{q_i\lfun{+_A}q_j\quad (q_i,m,n)\in s}{s\rightarrow s\setminus\set{(q_i,m,n)}\union\set{(q_j,m+1,n)}} &
\seq{q_i\lfun{-_A}q_j\quad (q_i,m,n)\in s\quad m>0}{s\rightarrow s\setminus\set{(q_i,m,n)}\union\set{(q_j,m-1,n)}}\\\\
\seq{q_i\lfun{+_B}q_j\quad (q_i,m,n)\in s}{s\rightarrow s\setminus\set{(q_i,m,n)}\union\set{(q_j,m,n+1)}} &
\seq{q_i\lfun{-_B}q_j\quad (q_i,m,n)\in s\quad n>0}{s\rightarrow s\setminus\set{(q_i,m,n)}\union\set{(q_j,m,n-1)}}
\end{array}$$
$$\seq{q_i\lfun{\it fork}q_j,q_k\quad (q_i,m,n)\in s}{s\rightarrow s\setminus\set{(q_i,m,n)}\union\set{(q_j,m,n),(q_k,m,n)}}$$
Note that the instructions $-_A$ and $-_B$ can only act on triplets of the form $(q,m,n)$ with $m>0$ and $n>0$, respectively. A triplet $(q,m,n)$ formalizes a machine in state $q$ whose two counters $A$ and $B$ hold values $m\geq 0$ and $n\geq 0$, respectively.
The last type of instructions is what makes these TCM's a variant of the standard ones, which have instead two zero-test instructions for the two counters $A,B$, and which act on triplets $(q,m,n)$, rather than on multisets of such triplets. The {\em fork} instruction necessitates
these more complicated instantaneous descriptions, which formalize a cluster of machines working in parallel, one for each triplet, or local ID, of the multiset.
A fork instruction executed by one of the machines M can be understood as launching a new TCM M' with initial local ID, say, $(q_k,m,n)$ while M evolves to local ID $(q_j,m,n)$. The counters are local to the machines, hence the counters of M'  can later evolve differently from the counters of M.

We are now ready for the encoding. States are encoded as (distinct) atomic formulas of the same name. Moreover, we  pick two fresh distinct atomic formulas $a$ and $b$, and we set
$a^0=1$, $a^1=a$, and $a^{n+1}=a^n\otimes a$, and similarly for $b$.  The fact that counter $A$ holds value $n$ will be encoded by the formula $a^n$. We next encode rules as follows:
$$\left.\begin{array}{r}
q_i\lfun{+_A}q_j\\
q_i\lfun{-_A}q_j\\
q_i\lfun{+_B}q_j\\
q_i\lfun{-_B}q_j\\
q_i\lfun{\it fork}q_j,q_k\\
\end{array}\right\}\mbox{ for }
\left\{\begin{array}{l}
q_i\vdash q_j\otimes a\\
q_i\otimes a\vdash q_j\\
q_i \vdash q_j\otimes b\\
q_i\otimes b\vdash q_j\\
q_i\vdash q_j\oplus q_k\\
\end{array}\right.$$
We have used bilateral sequents to help understanding:  one reads, say,  the first kind of rules as: \guil{exchange $q_i$ for $q_j$ and for one more $a$} (incrementing the counter $A$). We shall soon see why $\oplus$ is the right connective to use for the fifth sort of rules. Formally, we interpret each rule as a formula, as follows:
$$\begin{array}{l}
\dl q_i\lfun{+_A}q_j\dr= ?((q_i^\bot\lpar(q_j\otimes a))^\bot)\\
\dl q_i\lfun{-_A}q_j\dr = ?(((q_i\otimes a)^\bot\lpar q_j)^\bot)\\
\dl q_i\lfun{+_B}q_j\dr = ?((q_i^\bot\lpar(q_j\otimes b))^\bot)\\
\dl q_i\lfun{-_B}q_j\dr = ?(((q_i\otimes b)^\bot\lpar q_j)^\bot)\\
\dl q_i\lfun{\it fork}q_j,q_k\dr = ?((q_i^\bot\lpar( q_j\oplus q_k))^\bot)
\end{array}$$
The reason for the $?((\_)^\bot)$ format is that we want to be able to use the encodings of rules as
axioms, whenever needed, and hence possibly repetitively. We write $\vdash'\Gamma$ for
$\vdash \Gamma\union\dl I_1\dr\union\dl I_2\dr\union\dl I_3\dr\union\dl I_4\dr\union\dl I_5\dr$.
\begin{proposition}
Let M be a TCM, and $s$ be an ID for M. Then M accepts from $s$, i.e., there exists a sequence of transitions $s\rightarrow^\ast s'$ where $s'$ consists only of identical triplets $(q_F,0,0)$, if and only if, for every triplet $(q,m,n)\in s$, the sequent  $\vdash' q^\bot,(a^m)^\bot,(b^n)^\bot,q_F$ is provable.
\end{proposition}
Informally, the characterization in the statement reads as: assuming $q\otimes a^m\otimes b^n$ (the encoding of the
triplet $(q,m,n)$) and all the rules as axioms, then $q_F$ is provable, for all triplets of $s$.
Let us observe that $M$ accepts from $s$ if and only it accepts from each of the members of $s$,
since each triplet of $s'$ can be traced back to exactly one triplet of $s$, whence the ``for all'' form of the statement.

\medskip
\Proofhint
We first sketch the easy part of the proof, namely that accepting implies provable. The proof is by induction on the length of the reduction $s\rightarrow^\ast s'$. In the base case we have $s=s'$, hence
all the elements of $s$ are $(q_F,0,0)$, and the conclusion holds by weakening of the axiom
$\vdash q_F^\bot,q_F$. So, let
$s\rightarrow s_1\rightarrow^\ast s'$. If 
$s_1=s\setminus\set{(q_i,m,n)}\union\set{(q_j,m+1,n)}$, then all we are left to prove is that
$\vdash' q_i^\bot,(a^m)^\bot,(b^n)^\bot,q_F$, knowing that 
$\vdash' q_j^\bot,(a^{m+1})^\bot,(b^n)^\bot,q_F$. The proof is as follows:
$$\seq{\seq{\seqdbl{\seq{\vdots}{\vdash' q_j^\bot,(a^{m+1})^\bot,(b^n)^\bot,q_F}}{\vdash' q_j^\bot\lpar a^\bot,(a^m)^\bot,(b^n)^\bot,q_F}\quad\quad\seqdbl{\seq{\vdash (q_i^\bot\lpar(q_j\otimes a))^\bot,q_i^\bot\lpar(q_j\otimes a)}{\vdash ?((q_i^\bot\lpar(q_j\otimes a))^\bot),q_i^\bot\lpar(q_j\otimes a)}}{\vdash ?((q_i^\bot\lpar(q_j\otimes a))^\bot),q_i^\bot,q_j\otimes a}}
{\vdash' \:?((q_i^\bot\lpar(q_j\otimes a))^\bot),q_i^\bot,(a^m)^\bot,(b^n)^\bot,q_F}}
{\vdash' q_i^\bot,(a^m)^\bot,(b^n)^\bot,q_F}$$
(modulo the reversibility of $\lpar$, and using the contraction rule at the end).
We chek the case $s_1=s\setminus\set{(q_i,m,n)}\union\set{(q_j,m,n),(q_k,m,n)}$ more informally, as follows:
$$\seq{\seq{\seq{\vdots}{\vdash q_j^\bot,(a^m)^\bot,(b^n)^\bot,q_F}\quad\seq{\vdots}{\vdash q_k^\bot,(a^m)^\bot,(b^n)^\bot,q_F}}{\vdash q_j^\bot\& q_k^\bot,(a^m)^\bot,(b^n)^\bot,q_F}\quad\quad\seq{}{\vdash q_i^\bot,q_j\oplus q_k}}
{\vdash q_i^\bot,(a^m)^\bot,(b^n)^\bot,q_F}$$
Note that from a dialogue game point of view, the opponent of the proof has the choice
of which triplet to check, either $(q_j,m,n)$ or $(q_k,m,n)$, whence the presence of the
external choice connective $\&$ in the encoding $?(q_i\otimes(q_j^\bot\& q_k^\bot))$ of the fork rule. 

The converse direction (provable implies accepting) goes by analysing the cut-free proofs of
the encodings of acceptation, and by showing that they are essentially unique and that one can read back an accepting sequence of transitions. \qed

\medskip
Since  given $M$ and $s$ it is undecidable whether $M$ accepts from $s$, we conclude by reduction that LL is undecidable.

\begin{exercise} Formalize the proof of decidability of MALL outlined above. Hints: (1) associate with each sequent $\vdash\Delta$ a complexity measure $c(\vdash\Delta)$ such that for each instance of a MALL rule 
$$\seq{\mbox{antecedent 1}\;\;\ldots\;\;\mbox{antecedent }n}{\mbox{conclusion}}$$
we have $c(\mbox{antecedent i})<c(\mbox{conclusion})$ for all $i$; (2) then show that the set of \guil{possible proofs} of a sequent (i.e. the trees whose internal nodes correspond to correct application of the rules of MALL and whose leaves  are sequents $\Gamma$ where $\Gamma$ consists of atomic formulas only) can be effectively generated; (3) then show that one can effectively check which among these \guil{possible proofs} are real proofs (look at the leaves).
\end{exercise}

\section{Phase semantics} \label{phase-sem}

Phase semantics is a semantics of provability. Like in Kripke models for modal logics or intuitionisitic logic, a formula is not true or false, we rather have ways to express the extent to which it is true. In Kripke semantics, the meaning of a formula is the set of so-called worlds at which it holds. In phase semantics, a similar, but more structured, notion of {\em phases} is used.
Formulas are interpreted by {\em facts}, which are well-behaved sets of phases.

When setting up a notion of model, one has in mind both soundness and completeness of the class of models to be defined. This class must be large enough to include most natural examples
of candidate models, and in particular it must contain the candidate {\em syntactic model} which will serve to prove completeness. As a guide to the understanding of the notion of phase space, we expose
first what this syntactic model looks like. Its phases are just sequences of formulas. The intention is that the interpretation of a formula $A$ in this model is the set of all $\Gamma$'s such that
$\vdash \Gamma,A$. 
The soundness theorem will have the following form: if $\vdash A$, then $A$ is valid, in the following sense:
the special phase $1$ (think of it as expressing the certitude that $A$ holds) belongs to the interpretation of $A$. In the syntactic model, $1$ will be the empty sequence of formulas. Thus, in this model, the validity of $A$ is exactly the provability of $\vdash\emptyset,A$, i.e., of 
$\vdash A$. With these preparations, we can now define phase spaces. We deal with MALL first.

\begin{definition}
A {\em phase space} is a pair $(P,\bbot)$ where $P$ is a commutative monoid (in multiplicative notation: $pq$, $1$) and $\bbot\inc M$ is a distinguished subset of {\em antiphases}. One defines the following operation on subsets of $P$: $X^\bot=\setc{q}{\qqs{p\in X}{pq\in\bbot}}$.  
\end{definition}

It may be useful to think of the $q$'s as observers of the elements of $X$ and of $pq\in\bbot$ as a form of agreement between the
observer $q$ and the observed $p$ (somehow, a degenerated form of the kind of dialogue described in section \ref{mall}). We may say that ``$p$ passes the test $q$'' when $pq\in\bbot$.
A {\em fact} is a subset $F\inc P$ such that $F^{\bot\bot}=F$.

The following properties are easy to check, and the reader must have seen them somewhere already (linear algebra!):

$$\begin{array}{l}
X\inc Y \implies Y^\bot\inc X^\bot\\
X\inc X^{\bot\bot}\\
X^\bot=X^{\bot\bot\bot}\\
F\mbox{ is a fact if and only if }F=X^\bot\mbox{ for some }X\\
(X\union Y)^\bot =X^\bot\inter Y^\bot
\end{array}$$

\noindent
Thus a fact is a set of elements that pass the same tests, an idea which will be re-used in the richer setting of ludics. Notice also that facts obey $F^{\bot\bot}=F$, so that $^\bot$ is suitable for interpreting the linear negation, provided all formulas are interpreted by facts. Notice also that 
$\bbot=\set{1}^\bot$, hence $\bbot$ is a fact, which we use to interpret the constant $\bot$.
Also, $P=\emptyset^\bot$ is a fact.

We now define the various connectives as operations on facts. We write $XY=\setc{pq}{p\in X, q\in Y}$.

$$\begin{array}{lllll}
F\otimes G=(FG)^{\bot\bot} &&
\top=\emptyset^\bot &&
F\&G=F\inter G\;.
\end{array}$$ 

\noindent The rest can be deduced by duality:
$$F\lpar G=(F^\bot G^\bot)^\bot\quad F\oplus G=(F\union G)^{\bot\bot}\quad 0=P^\bot\quad 1=\bbot^\bot$$
(recall that $\bot$ is interpreted by $\bbot$). This determines the interpretation of  formulas, provided we have assigned facts to atomic formulas. We freely say that $A$ is interpreted by $A$, and if $\Gamma=A_1,\ldots,A_n$, we also write freely $\Gamma$ for (the interpretation of) $A_1\lpar\ldots\lpar A_n$.
The soundness theorem is stated as follows:
\begin{proposition} For any interpretation in a phase space, if  $\vdash A_1,\ldots,A_n$, then
$1\in A_1\lpar\ldots\lpar A_n$ (we say that $A_1,\ldots,A_n$ is valid).
\end{proposition}
\Proof We just check the case of $\bot$. 
Suppose $1\in \Gamma$. We have to show that $1\in (\Gamma^\bot\bbot^\bot)^\bot$, i.e., that
$\Gamma^\bot\bbot^\bot\inc\bbot$. Let $p\in\Gamma^\bot$ and $q\in\bbot^\bot$. Since $1\in\Gamma$, we have $p\in\bbot$, and since $q\in\bbot^\bot$, we have $pq\in\bbot$. \qed

\smallskip 
The following is a less symmetric, but more intuitive characterization of the notion of validity.
\begin{lemma} \label{1-inc}
For any $\Gamma=A_1,\ldots,A_n$, and any $1\leq i\leq n$, $\Gamma$ is valid if and only if
(the interpretation of) $A_i^\bot$
is included
in (the interpretation of) $A_1\lpar\ldots\lpar A_{i-1}\lpar A_{i+1}\lpar\ldots\lpar A_n$.
\end{lemma}
\Proof For simplicity, we take $n=2$,  and we set $A_1^\bot=A$ and $A_2=B$.
We have:
$$1\in (A^{\bot\bot}B^\bot)^\bot \iff A^{\bot\bot}B^\bot\inc\bbot\iff A^{\bot\bot}\inc B^{\bot\bot}\iff A\inc B\;. $$
\qed

\smallskip
Hence, if $A\approx B$, i.e., if $A\vdash B$ and $B\vdash A$, then $A$ and $B$ are interpreted by the same fact.

In particular, we have that $F\lpar (G\with H)=(F\lpar G)\with(F\lpar H)$, for arbitrary facts $F,G,H$. 
(take the formulas $P\lpar (Q\with R)$ and $(P\lpar Q)\with(P\lpar R)$, with $P,Q,R$ atomic and interpreted by $F,G,H$, respectively). As an exercise, we provide below an ad hoc proof of this equality, which makes use of the following property.
\begin{lemma} \label{XYbotbot}
Let $X,Y\inc P$. The following inclusion holds: $X^{\bot\bot}Y^{\bot\bot}\inc(XY)^{\bot\bot}$.
\end{lemma}
\Proof We have to show that $pqr\in\bot$ for any $p\in X^{\bot\bot}$, $q\in Y^{\bot\bot}$ and $r\in (XY)^\bot$. Taking some $f\in X$, we get that $rf\in Y^\bot$, hence $q(rf)=(rq)f\in\bbot$.
Since this was for arbitrary $f\in X$, we get $rq\in X^\bot$, and $p(rq)=pqr\in\bbot$. \qed

\smallskip\noindent
We have to prove $L =(F^\bot(G\inter H)^\bot)^\bot=(F^\bot G^\bot)^\bot\inter(F^\bot H^\bot)^\bot$.  The right hand side can be rewritten as follows: 
$$(F^\bot G^\bot)^\bot\inter(F^\bot H^\bot)^\bot = (F^\bot G^\bot\union F^\bot H^\bot)^\bot=
 (F^\bot (G^\bot\union H^\bot))^\bot=R\;.$$
Then we observe that $G\inter H\inc G$ entails $(G\inter H)^\bot\supseteq G^\bot$.
We have $(G\inter H)^\bot\supseteq H^\bot$, similarly. Hence $F^\bot(G\inter H)^\bot\supseteq
F^\bot(G^\bot\union H^\bot)$, from which $L\inc R$ follows. 
For the converse direction, we exploit the information that $F,G,H$ are facts, i.e., that 
$F=X^\bot$, $G=Y^\bot$, and $H=Z^\bot$ for some subsets $X,Y,Z$ of $P$. We have then, using lemma \ref{XYbotbot}:
$$\begin{array}{lll}
L =  (X^{\bot\bot}(Y^\bot\inter 
Z^\bot)^\bot)^\bot & =  &
(X^{\bot\bot}((Y\union Z)^{\bot\bot})^\bot\\
& \supseteq  & (X(Y\union Z))^{\bot\bot\bot}\\
 & = &
  (X(Y\union Z))^\bot\\
& \supseteq & (X^{\bot\bot}(Y^{\bot\bot}\union Z^{\bot\bot}))^\bot =  R\;.
\end{array}$$

\medskip We now define the syntactic model. As announced, $P$ is the set of the $\Gamma=A_1\ldots A_n$'s (up to the order of the formulas), written by juxtaposition (the monoid is given by concatenation), and
$\bbot=\setc{\Gamma}{\:\vdash\Gamma}$.

\begin{proposition} Phase spaces form a complete class of models, i.e., if  a MALL formula $A$ is valid in all phase spaces, then $\vdash A$.
\end{proposition}
\Proof
One shows by induction that the interpretation of $A$ in the syntactic model is
$\setc{\Gamma}{\:\vdash\Gamma,A}$, and we conclude then as indicated at the beginning of the section. Let us call ${\it Pr}(A)$ the latter set. For atomic formulas, we just fix the intepretation of $A$ to be ${\it Pr}(A)$, but to this aim we must first verify that
${\it Pr}(A)$ is a fact, which we prove for an arbitrary formula $A$. We have to show that if $\Delta\in
{\it Pr}(A)^{\bot\bot}$, then $\vdash\Delta,A$. It follows easily from the fact that $A\in{\it Pr}(A)^\bot$.
We check only that ${\it Pr}(A\otimes B)\inc({\it Pr}(A){\it Pr}(B))^{\bot\bot}$. Let
$X={\it Pr}(A){\it Pr}(B)=\setc{\Gamma_1\Gamma_2}
{\:\vdash A,\Gamma_1\mbox{ and } \vdash B,\Gamma_2}$. We have to prove that
if  $\Delta\in X^\bot$, then, for all $\Gamma$,
$\vdash A\otimes B,\Gamma$ entails $\vdash \Gamma,\Delta$. 
We notice that $A^\bot B^\bot\in X$ since $\vdash A^\bot,A$ and $\vdash B^\bot,B$. Hence
$\vdash A^\bot,B^\bot,\Delta$, and:
$$\seq{\vdash A\otimes B,\Gamma \quad\seq{\vdash A^\bot,B^\bot,\Delta }{\vdash A^\bot\lpar B^\bot,\Delta}}{\vdash \Gamma,\Delta}$$
\qed

\medskip
We next give the phase semantics of exponentials. Like for intuitionistic logic, topological ideas 
are useful. In the case of intuitionistic logic, in order to avoid the tautology ($A$ or $\neg A$), one interprets, say, $\neg A$ by the interior of the complement of $A$ in some topology. In the case of linear logic, we still have to choose the right infinite conjunction and the finite disjunction used to define the ``closed'' sets.

\begin{definition} A {\em topolinear} space is a structure $(P,\bbot,{\bf F})$, where $(P,\bbot)$ is a phase space, and where {\bf F} is a set of facts that satisfies the following conditions:
$$\begin{array}{ll}
(1)& \mbox{ {\bf F} is closed by arbitrary intersections},\\
(2)& \mbox{ {\bf F} is closed by (finite) } \lpar,\mbox{ and (in particular) }\bbot\in{\bf F},\\
(3)& \bbot \mbox{ is the smallest fact of  {\bf F}},\\
(4)& F\lpar F\in{\bf F},\mbox{ for all }F\in{\bf F}.
\end{array}$$

\noindent
The facts of  {\bf F} are called the {\em closed} facts.
Given a fact $F$, we define $?F$ as the smallest 
closed fact containing $F$.
\end{definition}
Notice that, defining the open facts as the orthogonals of closed facts, we get a definition of $!F$ as the largest open fact contained in $F$. 

\smallskip
The soundness theorem extends to full LL.  For weakening, we proceed exactly as for the $\bot$ rule, using $\bbot\inc ?A$.  The validity of contraction is an obvious consequence of axiom (4).  Using lemma 
\ref{1-inc}, the validity of dereliction and of promotion are immediate consequences of the axioms
(3) and (2), respectively.

\smallskip
We  extend the definition of the syntactic model as follows.
We take as closed facts arbitrary intersections of sets of the
form ${\it Pr}(?A)$. Let us verify that this collection is closed under finite $\lpar$'s.
We have proved $F\lpar (G\with H)=(F\lpar G)\with(F\lpar H)$ above. The infinitary version of this equality is proved in the same way, so that we get
$$\begin{array}{lll}
(\Inter_{i\in I}{\it Pr}(?A_i))\lpar(\Inter_{j\in J}{\it Pr}(?B_j)) & = &
\Inter_{i\in I,j\in J}({\it Pr}(?A_i)\lpar{\it Pr}(?B_j))\\
& = &
\Inter_{i\in I,j\in J}{\it Pr}(?A_i\lpar?B_j)\\
& = &
\Inter_{i\in I,j\in J}{\it Pr}(?(A_i\oplus B_j))
\end{array}$$
where the last equality follows from  $?(A_i)\lpar?(B_j)\approx?(A_i\oplus B_j)$. 
Also, $\bbot$ is a closed fact since $\bbot\approx\; ?0$, and it is the smallest one, by the weakening rule. Axiom (4) is the consequence of two observations: 
$$\begin{array}{lllllll}
(\dagger) & ?A\approx ?A\lpar?A &&&&
(\ddagger) & ?A_i\vdash ?A_i,?A_j\;.
\end{array}$$

\noindent
$(\dagger)$ follows from $A\vdash A\oplus A$ (by either of the $\oplus$ rules) and from $\vdash A^\bot\& A^\bot, A$), and $(\ddagger)$ is immediate by weakening. Then we have:
$$\begin{array}{lll}
(\Inter_{i\in I}{\it Pr}(?A_i))\lpar(\Inter_{j\in J}{\it Pr}(?A_j)) & = &
\Inter_{i\in I,j\in J}({\it Pr}(?A_i)\lpar{\it Pr}(?A_j))\\
& = &
\Inter_{i\in I}({\it Pr}(?A_i)\lpar{\it Pr}(?A_i))\quad \mbox{by }(\ddagger)\\
& = &
\Inter_{i\in I}{\it Pr}(?A_i)\quad \mbox{by }(\dagger)
\end{array}$$
Thus the syntactic structure is a topolinear space. 

\smallskip
 We have to prove that ${\it Pr}(?A)=\Inter\setc{{\it Pr}(?B)}{{\it Pr}(A)\inc {\it Pr}(?B)}$. Since ${\it Pr}(A)\inc {\it Pr}(?A)$ by dereliction,
we have one inclusion. For the other, we have to show that if ${\it Pr}(A)\inc {\it Pr}(?B)$ then 
${\it Pr}(?A)\inc {\it Pr}(?B)$. So let $\vdash \Gamma,?A$. Since ${\it Pr}(A)\inc {\it Pr}(?B)$, then in particular
$\vdash A^\bot,?B$, hence $\vdash !A^\bot,?B$ by promotion. We conclude that $\Gamma\in {\it Pr}(?B)$ by using the cut rule. This completes the proof of completeness.

\section{Coherence spaces} \label{coh-spaces}

With coherence spaces, we give a semantics of proofs. Coherence spaces may be understood as concrete descriptions of certain sets or {\em domains}, in the terminology of denotational semantics. They are a simplified version of event structures \cite{Winskel86}.

\begin{definition}
A {\em coherence space}\index{coherence space} $(E,\coh)$ ($E$ for 
short\index{$\coh$}) is given by a 
set $E$ of events, or tokens, and by a binary reflexive and symmetric 
relation $\coh$ over $E$. 
$E$ is called the {\em web}\index{web} of $(E,\coh)$. A state, or {\em clique}, of $E$ is a set $x$ of 
tokens satisfying the following consistency
condition:
$$\qqs{e_1,e_2\in x}{e_1\coh e_2}.$$
We denote with $D(E)$ the set of states of $E$, ordered by inclusion.
If $(E,\coh)$ is a coherence space, its {\em 
incoherence}\index{incoherence} is the relation defined by:
$$e_1\icoh e_2\Leftrightarrow \neg(e_1\coh e_2)\mor e_1=e_2.$$
Notice that the incoherence
is not the complement of the coherence, since the coherence and the
incoherence are both reflexive.  
We also define {\em strict incoherence} and {\em strict coherence} as follows, respectively:
$$e_1\smile e_2 \Leftrightarrow \neg(e_1\coh e_2)\quad\quad
e_1\frown e_2  \Leftrightarrow \neg(e_1\icoh e_2)$$
\end{definition}
Clearly, coherence can be recovered from incoherence:
$$e_1\coh e_2\Leftrightarrow \neg(e_1\icoh e_2)\mor e_1=e_2.$$
In fact, it is easy to check that all the relations $\coh$, $\icoh$, $\frown$ and $\smile$ are interdefinable, that is, if one of them is given, then the other three can be defined from it. 

\medskip
We illustrate coherence spaces with a few ewamples. The coherence space {\bf Nat}=$(\omega,\coh)$, where $m\coh n$ if and only if $m=n$ is such that
$D({\bf Nat})=\set{\emptyset}\union\setc{\set{n}}{n\in\omega}$, that is, $D({\bf Nat})$ is
isomorphic to the partial order $\set{\bot}\union\omega$, ordered by the {\em flat} ordering
$x\leq y$ if and only if $x=\bot$ or $x=y$. This partial order is used in denotational semantics to 
interpret the type of natural numbers. 

\medskip
We next show how to interpret the connectives of linear logic as constructions of coherence spaces.

\begin{definition}
The product, or ``with'',  $E\with E'$ of 
two  coherence spaces $E$ and $E'$  
is the coherence space whose tokens are  either $e.1$, with $e\in 
E$, or $e'.2$, with $e'\in E'$ 
(using an explicit notation for disjoint unions), and the coherence is given by:
$$(e_1.i)\coh(e_2.j)\Leftrightarrow i\neq j \mor (i=j\mand e_1\coh e_2).$$
\end{definition}

This definition is such that $D(E\& E')$ is isomorphic to $D(E)\times D(E')$, where $\times$ is the good old set-theoretical cartesian product. For example, taking $E=E'={\bf Nat}$, we have
that $(\bot,\bot)$, $(m,\bot)$, $(\bot,n)$, and $(m,n)$ are represented as
$\emptyset$, $\set{m.1}$, $\set{n.2}$, and $\set{m.1,n.2}$, respectively.

\begin{definition}
The tensor product  $E\otimes E'$ of 
two  coherence spaces $E$ and $E'$  
is the coherence space whose tokens are pairs $(e,e')$ where $e\in E$ 
and $e'\in E'$, and whose coherence is given by:
$$(e_1,e'_1)\coh(e_2,e'_2)\Leftrightarrow (e_1\coh e_2\mand e'_1\coh e'_2).$$
\end{definition}

We next introduce the duality, or linear negation.

\begin{definition}
The linear negation\index{linear 
negation} $E^{\bot}$\index{$E^{\bot}$} of a 
coherence space
$(E,\coh)$ is defined as
$E^{\bot}=(E,\icoh)$.
\end{definition}

The definition of the interpretation of $\lpar$ and $\limpl$ (and also of $\oplus$ -- left to the reader) is  then obtained by De Morgan duality.

\begin{definition}
Let $E$, $E'$ be coherence spaces. Their ``par'' $E\lpar E'$\index{$E\lpar E'$}  is the coherence space whose tokens are pairs 
$(e,e')$ where $e\in E$ and $e'\in E'$, and whose incoherence is given 
by:
$$(e_1,e'_1)\icoh(e_2,e'_2)\Leftrightarrow (e_1\icoh e_2\mand e'_1\icoh 
e'_2).$$
\end{definition}
\begin{definition}
Let $E$, $E'$ be coherence spaces. Their linear 
implication $E\limpl E'$\index{$E\lpar E'$} is the coherence space whose tokens are pairs 
$(e,e')$ where $e\in E$ and $e'\in E'$, and whose incoherence is given 
by:
$$(e_1,e'_1)\icoh(e_2,e'_2)\Leftrightarrow (e_1\coh e_2\mand e'_1\icoh 
e'_2).$$
\end{definition}

\begin{exercise} \label{impl-coh}
Show that the following are other equivalent definitions of the coherence of the linear function space:
$$\begin{array}{ll}
(1) & (e_1,e'_1)\coh(e_2,e'_2)\Leftrightarrow (e_1\coh e_2\Rightarrow
(e'_1\coh e'_2\mand (e_1\neq e_2\Rightarrow e'_1\neq e'_2)))\\
(2) & (e_1,e'_1)\coh(e_2,e'_2)  \Leftrightarrow  (e_1\coh e_2\Rightarrow 
e'_1\coh e'_2)\mand
(e'_1\icoh e'_2\Rightarrow e_1\icoh e_2)\;.
\end{array}$$ 

\end{exercise}

\begin{definition}
Let $(E,\coh)$ be a coherence space.
The exponential $!E$ 
is the coherence space whose 
tokens are the finite cliques of $E$, and whose coherence is given by
$(x_1\coh x_2\Leftrightarrow x_1\uparrow x_2)$, where $x_1\uparrow x_2$ means that there exists a clique $x$ of $E$ such that $x_1\inc x$ and $x_2\inc x$.
\end{definition}

Now we explain briefly the interpretation of LL in coherence spaces.
\begin{enumerate}
\item A formula $A$ is interpreted by a coherence space (it does not harm to use the same name, as we did already for phase semantics).
 \item A proof $\pi$ of a sequent $\vdash A_1,\ldots,A_n$ is interpreted as a clique $\dl\pi\dr$ of $A_1\lpar\ldots\lpar A_n$. 
\item If $\pi$ rewrites through cut-elimination to $\pi'$, then  $\dl\pi\dr=\dl \pi'\dr$.
\end{enumerate}
As an example, we show (2) for the cut rule. We first remark that a token of $A_1\lpar\ldots\lpar A_n$ is a vector $\vec{e}=(e_1,\ldots,e_n)$, and that (cf. the definition of $\lpar$) we have
$$(\dagger)\quad \vec{e}\frown\vec{f}\quad \mbox{if and only if}\quad e_i\frown f_i\mbox{ for some }i\;.$$
Let $\pi_1$ be a proof of $\vdash\Gamma, A$ and $\pi_2$ be a proof of 
$\vdash\Gamma',A^\bot$. and let  $\pi$ be the proof of $\vdash\Gamma,\Gamma'$ resulting from applying the cut rule. We define the interpretation of $\pi$ as follows  (relation composition!):
$$\dl\pi\dr=\setc{(\vec{e},\vec{e'})}{\xst{a}{(\vec{e},a)\in\dl\pi_1\dr\;\mbox{and}\;(\vec{e'},a)\in\dl\pi_2\dr}}\;.$$
We have to show that $\dl\pi\dr$ is a clique. Let $(\vec{e},\vec{e'}),(\vec{f},\vec{f'})\in\dl\pi\dr$.
Then there exist $a$ and $b$ such that
$(\vec{e},a)\in\dl\pi_1\dr, (\vec{e'},a)\in\dl\pi_2\dr, (\vec{f},b)\in\dl\pi_1\dr$, and  $(\vec{f'},b)\in\dl\pi_2\dr$. So, because we know by induction that $\dl\pi_1\dr$ and
$\dl\pi_2\dr$ are cliques, we have $(\vec{e},a)\coh(\vec{f},b)$ and $(\vec{e'},a)\coh(\vec{f'},b)$.
If both $(\vec{e},a)=(\vec{f},b)$ and $(\vec{e'},a)=(\vec{f'},b)$, then we have also 
$(\vec{e},\vec{e'})=(\vec{f},\vec{f'})$ and a fortiori $(\vec{e},\vec{e'})\coh(\vec{f},\vec{f'})$, so we are done. Let us suppose thus that, say, $(\vec{e},a)\frown(\vec{f},b)$. By property $(\dagger)$, there are two cases: either we have $e_i\frown f_i$ for some $i$, or $a\frown_A b$ (the subscript $A$ means that $\frown$ is relative to $A$). In the first case, we have
directly that $(\vec{e},\vec{e'})\frown(\vec{f},\vec{f'})$, using $(\dagger)$. In the second case, we know that $a\neq b$ (by the definition of $\frown$), so a fortiori we have 
$(\vec{e'},a)\frown(\vec{f'},b)$. Hence, by $(\dagger)$, either there exists $i'$ such that
$e'_{i'}\frown f'_{i'}$, and then we conclude as above, or $a\frown_{A^\bot} b$. But this cannot happen, because we have $a\frown_A b$, hence a fortiori $a\coh_A b$, which can be rephrased as
$a\icoh_{A^\bot} b$ and in turn can be rephrased as $\neg(a\frown_{A^\bot}b)$. This ends the proof.

\medskip But there is more to this semantics. Coherence spaces can be organized in a category, actually two categories: the category of stable functions (this is where it all started), and the category of linear functions. The domains $D(E)$ are closed under union of increasing sequences, that is, they are {\em complete partial orders}. It remains to find appropriate morphisms between coherence spaces, defined as appropriate functions from cliques to cliques.

\begin{definition} Let $(E,\coh)$ and $(E',\coh)$ be two coherence spaces. A mono\-tonous function
$f:D(E)\rightarrow D(E')$ is called {\em stable} if it is continuous, i.e., $f(\Union_{n\in\omega} x_n)=\Union_{n\in\omega} f(x_n)$ for every increasing sequence $x_n$, and if it preserves compatible intersections, i.e.:
$$\qqs{x,y}{x\uparrow y\implies f(x\inter y)=f(x)\inter f(y)}$$
(notice that if $x\uparrow y$, then $x\inter y$ is a clique).
If moreover $f$ preserves compatible unions, i.e.:
$$\qqs{x,y}{x\uparrow y\implies f(x\union y)=f(x)\union f(y)}$$
then $f$ is called {\em linear}.
Stable functions are ordered by the {\em stable ordering} $\leq_s$,  defined as follows:
$f\leq_s g$ if and only if $(\qqs{x\in D(E)}{f(x)\leq g(x)})$ (pointwise ordering) and
$$\qqs{x\leq y}{f(x)=g(x)\inter f(y)}\;.$$ 
\end{definition}

\begin{proposition}  \label{birth-LL}
For all $(E,\coh)$, $(E',\coh)$, $({\bf Coh}[E,E'],\leq_s)$ (the partial order of stable functions) is order-isomorphic to $(D(!E\multimap E'),\inc)$.
\end{proposition}
\Proofhint
The inverse bijections are defined as follows. Given a stable function $f$ we define its {\em trace}, and given a clique $\phi$ of $!E\multimap E'$ we define the inverse transformation as follows, respectively:
$$\begin{array}{l}
{\it trace}(f)=\setc{(x,e')}{e'\in f(x)\mbox{ and } (\qqs{y<x}{e'\nin f(y)})}\\
{\it fun}(\phi)(z)=\setc{e'}{\xst{x}{x\leq z\mbox{ and }(x,e')\in\phi}}\;.
\end{array}$$
The most interesting part of the proof consists in verifying that ${\it trace}(f)$ is a clique, and that the stable ordering is just the inclusion of traces. One uses the following characterization of
the coherence relation of $!E\multimap E'$  (cf. exercise  \ref{impl-coh}):
$$ (x_1,e'_1)\coh(x_2,e'_2)\Leftrightarrow (x_1\coh x_2\Rightarrow
(e'_1\coh e'_2\mand (e'_1=  e'_2\Rightarrow x_1= x_2)))\;.$$
The trace has an operational flavour: one can read $(x,e')\in{\it trace}(f)$ and $x\leq z$ as: ``$x$ is {\em the} (finite) part of $z$ which is used to compute the output event $e'\in f(z)$''. Indeed, $x$ is unique with that property, by the above characterization of the coherence, and is called the minimum point for $y$ and $e'$
The stable ordering now also has an operational flavour: if $f\leq_s g$, then ``$f$ and $g$ are computed in the same way'' in the sense that $g$ has to respect the minimum points of $f$. We show that $f\leq_s g$ implies ${\it trace}(f)\inc{\it trace}(g)$. Let $(x,e')\in{\it trace}(f)$. Then $e'\in f(x)\inc g(x)$, and hence
there exists $(y,e')\in{\it trace}(g)$ such that $y\leq x$. We show that $y=x$, which will establish the inclusion. Suppose $z<x$. Then we would have $e'\in f(y)=f(x)\inter g(y)$, which contradicts the minimality of $x$. The converse implication is proved similarly.
\qed

\medskip
This proposition has a capital historical importance: it is the observation of the quasi-symmetry of
input and output in the traces of stable functions which lead Girard to force a complete symmetry
by deciding that finite cliques can be considered as atoms, or events of a new coherence space,
and hence by decomposing $E\rightarrow E'$ as $!E\multimap E'$.
\begin{exercise} \label{graph-linfun}
Show that for all $(E,\coh)$, $(E',\coh)$, $({\bf Coh_l}[E,E'],\leq_s)$ (the partial order of linear functions) is order-isomorphic to $(D(E\multimap E'),\inc)$.
\end{exercise}

\section{Categorical models} \label{cat-mod}

In this section, we describe a general categorical semantics for linear logic, and we introduce the
appropriate categorical apparatus. We assume that the reader knows about categories, functors, natural transformations, and adjunctions. We briefly reintroduce other rather standard notions such as monoidal categories and comonads, as well as more ad hoc notions and axioms that are needed to complete the picture.

\begin{definition} \label{mon-cat-Linear}
A {\em monoidal} category\index{monoidal category} is a category \cat{C} 
equipped with a functor $\otimes:\cat{C}\times\cat{C}\rightarrow\cat{C}$, called 
the {\em tensor product}\index{tensor product},
 a distinguished object $1$, called the {\em tensor 
unit}\index{tensor unit}, and
natural isomorphisms, also called the canonical isomorphisms:
$$\begin{array}{lllll}
\alpha:A\otimes(B\otimes C)\rightarrow(A\otimes B)\otimes C &&
\iota_l:1\otimes A\rightarrow A &&
\iota_r:A\otimes 1\rightarrow A\; ,
\end{array}$$
satisfying the following two so-called coherence equations:
$$\begin{array}{lllll}
(\alpha-\alpha) & \alpha\comp\alpha = 
(\alpha\otimes\id)\comp\alpha\comp(\id\otimes\alpha) &&
(\alpha-\iota) & (\iota_r\otimes\id)\comp\alpha = 
\id\otimes\iota_l\; .
\end{array}$$
\end{definition}
Where do the two coherence equations come from? As observed by Huet (unpublished), a good answer comes from 
rewriting theory (a subject that did not exist as such when monoidal categories 
were defined by Mac Lane in the early sixties). Consider the domains and 
codomains of the canonical 
isomorphisms  as the left and right hand sides of rewriting rules 
and rewriting sequences, respectively:
$$\begin{array}{lllllllll}
(\alpha) \;\; A\otimes(B\otimes C)  \rightarrow  (A\otimes B)\otimes C &&
(\iota_l) \;\; 1\otimes A  \rightarrow  A &&
(\iota_r) \;\; A\otimes 1  \rightarrow A
\end{array}$$

\noindent
The two coherence 
equations correspond to equating different reduction sequences:
$\alpha\comp\alpha$ encodes
$$
A\otimes(B\otimes (C\otimes D)) \rightarrow (A\otimes B)\otimes(C\otimes 
D)\rightarrow ((A\otimes B)\otimes 
C)\otimes D\;,$$
while $(\alpha\otimes\id)\comp\alpha\comp(\id\otimes\alpha)$ encodes
$$A\otimes(B\otimes (C\otimes D)) \rightarrow A\otimes ((B\otimes C)\otimes 
D)\funt  ((A\otimes B)\otimes C)\otimes D\;.$$
Similarly, the two sides of the second equation encode
$$\begin{array}{l}
A\otimes (1\otimes B)\rightarrow(A\otimes 1)\otimes B\rightarrow A\otimes B\\
A\otimes (1\otimes B)\rightarrow A\otimes B\; .
\end{array}$$
These pairs of derivations form local confluence 
diagrams for the rewriting system 
on objects induced by $(\alpha),(\iota_l),\mand(\iota_r)$. We pursue this 
interpretation in exercise 
\ref{mon-KB-Linear}, which assumes basic familiarity with
rewriting theory.
\begin{exercise} \label{mon-KB-Linear}
(1) Find all the critical pairs of the rewriting system on objects underlying  
$\alpha,\iota_l,\mand\iota_r$, and show that the corresponding 
equations between canonical 
isomorphisms are derivable from the
two equations given in definition \ref{mon-cat-Linear}.
Hint: There are three other critical pairs; exploit the fact 
that $\alpha,\iota_l,\mand\iota_r$ are 
isos.
(2) Prove the so-called coherence theorem for monoidal categories: 
every two canonical morphisms (that is, terms over the signature
$\{\comp$, $\id$, $\otimes$, $\alpha$, $\alpha^{-1}$, $\iota_l$, $\iota_l^{-1}$, $\iota_r$, $\iota_r^{-1}\}$) 
with the same domain and codomain are equal.
Hint: Remove first  $\alpha^{-1},\iota_l^{-1},\mand\iota_r^{-1}$, 
and proceed as in the proof of 
Knuth-Bendix theorem (confluence of critical pairs implies local 
confluence) \cite{Hu-Op-80}.
\end{exercise}
\begin{definition} \label{sym-mon-Linear}
A symmetric monoidal category\index{symmetric monoidal category} is a 
monoidal category together with an additional canonical isomorphism 
$\gamma: A\otimes B\rightarrow B\otimes A$
satisfying:
$$\begin{array}{llll}
(\gamma-\gamma) & \gamma\comp\gamma & = & \id\\
(\alpha-\gamma) & \alpha\comp\gamma\comp\alpha & = & 
(\gamma\otimes\id)\comp\alpha\comp(\id\otimes\gamma)\; .
\end{array}$$
\end{definition}
The coherence theorem still holds in the symmetric monoidal 
case, but needs more care: clearly we do not want to identify $\gamma:A\otimes A\rightarrow 
A\otimes A$ and 
$\id:A\otimes A\rightarrow A\otimes A$. 
Category theorists exclude this by speaking, not of terms, but of 
natural transformations: in the present case, we see that
$\gamma:(\lbd(A,B).A\otimes B)\rightarrow(\lbd(A,B).B\otimes A)$ and 
$\id:(\lbd(A,B).A\otimes B)\rightarrow(\lbd(A,B).A\otimes B)$
do not have the same codomain. A more elementary point of view is to 
restrict attention to 
linear terms for objects. 
\begin{exercise}
Show that, in a symmetric monoidal category, any two canonical 
natural transformations between the same functors are equal.
Hints:
Use monoidal coherence, and the following presentation of 
the symmetric group by means of the transpositions $\sigma_i$ which 
permute two 
successive elements $i$ and $i+1$:
$$\begin{array}{lll}
\sigma_i\comp\sigma_i=\id & 
\sigma_i\comp\sigma_j=\sigma_j\comp\sigma_i
\;\;(j-i>1) &
\sigma_i\comp\sigma_{i+1}\comp\sigma_i=
\sigma_{i+1}\comp\sigma_i\comp\sigma_{i+1} .
\end{array}$$
\end{exercise}
\begin{definition}
A monoidal closed category\index{monoidal closed category} is a 
monoi\-dal
category \cat{C} such that for all $A$ the functor $\lbd 
C.(C\otimes A)$
has a right adjoint, written $\lbd B.(A\limpl B)$. In other words, 
for every objects $A,B$, there exists an object $A\limpl B$, called the
{\em linear exponent}\index{linear exponent}, and natural bijections 
(for all $C$):
$$\Lambda_l:\cat{C}[C\otimes A,B]\rightarrow\cat{C}[C,A\limpl B].$$
\end{definition}

Notice that there are no accompanying additional coherence equations 
for monoidal closed categories. This comes from the difference in nature 
between the
constructions $\otimes$ and $\limpl$: the latter
is given together with a universal construction (an adjunction), 
while 
the first is just a functor with
some associated isomorphisms. This difference is often referred to as 
the difference between ``additional structure'' $(\otimes)$ and 
``property''
$(\limpl)$. The notion of dualizing object, due to Barr \cite{Barr91} and introduced next, is 
additional structure.
\begin{definition}
A symmetric monoidal closed category \cat{C} is called 
$\ast$-autonomous\index{$\ast$-autonomous} if
it has a distinguished object $\bot$, called a {\em dualizing 
object}\index{dualizing object},
such that for any $A$ the morphisms  
$$\Lambda_l(\Lambda_l^{-1}(\id)\comp\gamma):\cat{C}[A,(A\limpl\bot)\limpl\bot],$$ 
called canonical, have an inverse.
We write $A^{\bot}$ for $A\limpl\bot$, and 
$A^{\bot\bot}$ for $(A^{\bot})^{\bot}$.
\end{definition}
\begin{exercise} \label{closed-dual-Linear}
Let \cat{C} be a  $\ast$-autonomous category. Show that:
(1) there exists a natural bijection between 
$\cat{C}[A,B]$ and
$\cat{C}[B^{\bot},A^{\bot}]$;
(2) there exists a natural  isomorphism $(A\limpl 
B)^{\bot}\cong 
A\otimes B^{\bot}$; (3)
there exists a natural  isomorphism $1\cong \bot^{\bot}.$
\end{exercise}
Part (3) of exercise \ref{closed-dual-Linear} shows in 
retrospective that the name $\bot$ for the dualizing object is 
deserved: we can indeed understand it as the  multiplicative false.
\begin{exercise} \label{dual-funct-Linear}
Suppose that \cat{C} is a symmetric monoidal category, and that 
$(\_)^{\bot}:\cat{C}^{\it op}\rightarrow\cat{C}$ is a functor given together 
with: (1) a natural isomorphism $A\cong A^{\bot\bot}$;
(2) a natural bijection $\cat{C}[I,(A\otimes B^{\bot})^{\bot}] 
\cong 
\cat{C}[A,B]$.
Show that $\cat{C}$ is monoidal closed, with $\limpl$ defined by 
$A\limpl B = (A\otimes B^{\bot})^{\bot}$.
\end{exercise}

\medskip
The last ingredient we need is the notion of comonad.
\begin{definition}
A comonad\index{comonad} over a category $\cat{C}$ is a triple 
$(T, \epsilon, \delta)$ where $T: \cat{C} \rightarrow \cat{C}$ is a
functor, $\epsilon: T \rightarrow id_{\cat{C}}$, 
$\delta: T \rightarrow T\comp T$
are natural transformations, and the
following equations hold:
\[\begin{array}{lllll}
\epsilon_{TA}\comp\delta_A=\id_{TA}
&& T\epsilon_A\comp\delta_A=\id_{TA} 
&& \delta_{TA}\comp\delta_A=T\delta_A\comp\delta_A\;,
\end{array}\]
where e.g. $\delta_A:TA\rightarrow T(TA)$ stands for the component of $\delta$ at $A$.
The following derived operation is useful. For all
$f:TA\rightarrow B$, one constructs $\kleisli{f}:TA\rightarrow TB$ as follows:
$$\kleisli{f}=Tf\comp\delta.$$
We define the co-Kleisli category $\cat{C}_T$ as follows. The objects of $\cat{C}_T$ are the 
objects 
of $\cat{C}$, and for any $A,B$:
$$\cat{C}_T[A,B]=\cat{C}[TA,B].$$
The identity morphisms are given by $\epsilon$, and composition 
$\comp_T$
is defined by:
$$g\comp_T f= g\comp\kleisli{f}.$$
\end{definition}

Comonads are tightly linked with adjunctions, as the following exercises evidentiate.

\begin{exercise} \label{adj-com-Linear}
Show that every adjunction $(F,G,\eta,\epsilon)$, where $F:\cat{C}\rightarrow\cat{C'}$ 
and $G:\cat{C'}\rightarrow\cat{C}$, induces a comonad 
$(F\comp G,\epsilon,\delta)$ on $\cat{C}'$, where $\epsilon$ is the 
counit of the 
adjunction, and where $\delta=F\eta G$, i.e.,  
$\delta_B=F(\eta_{GB})$ (for all $B$).
Show that the co-Kleisli category associated with
the comonad is equivalent to the full subcategory of \cat{C} whose 
objects are
in the image of $G$.
\end{exercise}
\begin{exercise} \label{comon-adj}
Let $(T, \epsilon, \delta)$ be a comonad over a category $\cat{C}$. Show that the following data
define adjoint functors $F$ and $U$ between \cat{C} and $\cat{C}_T$ in such a way that $T=FU$:
$$\begin{array}{llllll}
FA=\:!A & Ff=\kappa(f) &&&
{\it UA}=A & Ug=g\comp\epsilon
\end{array}$$
with $\id$ as unit and $\epsilon$ as counit of the adjunction.
\end{exercise}

We next follow  \cite{Seely89} for a first attempt of interpretation of LL. We recall that a category \cat{C}  is cartesian if it has a terminal object $\top$ (i.e., such that for any $A$ the homset $\cat{C}[A,\top]$ has exactly one arrow $\top_A$) and binary products (i.e., for every objects $A,B$ there exists $(A\& B,\pi:A\& B\rightarrow A,\pi':A\& B\rightarrow B)$ such that for all $C$,
$f:C\rightarrow A$, and $g:C\rightarrow B$ there exists a unique arrow $\pair{f}{g}:C\rightarrow A\& B$ such that $\pi\comp\pair{f}{g}=f$ and $\pi'\comp\pair{f}{g}=g$; the product is then a functor, with $f\&g=\pair{f\comp\pi}{f\comp\pi'}$).
\begin{definition} \label{Seely-cat-Linear}
A {\em Seely category}
is a structure consisting of 
the following data:
(1) a $\ast$-autonomous category $\cat{C}$ which is at the same 
time cartesian; (2) a comonad $(!,\epsilon,\delta)$ over $\cat{C}$, called the 
{\em exponential}\index{exponential}, together
with two natural isomorphisms:
$$\begin{array}{lll}
n_{A,B}:  (!A)\otimes(!B) \cong !(A\& B)  & &
p: 1   \cong !\top \; .
\end{array}$$
\end{definition}
\begin{proposition} \label{implied-CC-Linear}
If $\cat{C}$ is a Seely category, then
the associated co-Kleisli category
$\cat{C}_!$  is cartesian.
\end{proposition}
\Proof We take the 
product on objects and the pairing of arrows of $\cat{C}$.
As projections we take
$\pi_{1}\comp\epsilon$ and $\pi_{2}\comp\epsilon.$
We check one commutation diagram:
$$\begin{array}{lllll}
(\pi_{1}\comp\epsilon)\comp_!\pair{f}{f'} & = & 
\pi_{1}\comp(\epsilon\comp\kleisli{\pair{f}{f'}})&&\\
& = & \pi_{1}\comp\pair{f}{f'} & = & f\; .
\end{array}$$\qedm
\begin{exercise} \label{implied-CCC-Linear}
Show that if $\cat{C}$ is a Seely category, then
the associated co-Kleisli category
$\cat{C}_!$  is cartesian closed (i.e., for any object $A$ the functor $\lbd 
C.(C\& A)$
has a right adjoint).
\end{exercise}
\begin{exercise} \label{CCC-tensor-bang-Linear}
Let $\cat{C}$ be a $\ast$-autonomous category which is at the same 
time cartesian, and which is equipped with a comonad 
$(!,\epsilon,\delta)$ 
such that the associated Kleisli category $\cat{C}_!$  is cartesian closed. Show that 
there
exists a natural isomorphism from $(!A)\otimes(!B)$ to
$!(A\with B)$.
\end{exercise}

Another implied structure is that each object of the form $!A$ is 
endowed with the structure of a commutative comonoid: there 
are two arrows
$$\begin{array}{lll}
e:\: !A\rightarrow 1 & & d:\: !A\rightarrow(!A)\otimes(!A)
\end{array}$$
satisfying 
the three (categorical versions of the) comonoid laws (see exercise
\ref{comonoid-laws-Linear}). These arrows are constructed as follows:
$$\begin{array}{lll}
e = p^{-1}\comp !(\top) \comp\delta &&
d  =n^{-1}\comp\:!(\pair{\epsilon}{\epsilon})\comp\delta\;.
\end{array}$$
This definition may seem ad hoc, but it is actually derived from the
underlying co-Kleisli adjunction (exercise \ref{comon-adj}) and from the comonoid structure induced by the cartesian products of $\cat{C}_!$ (proposition \ref{implied-CC-Linear} and exercise \ref{prod-com}): $e=p^{-1}\comp F\top$ and $d=n^{-1}\comp F\Delta$.
\begin{exercise} \label{prod-com}
Let $\cat{C}$ be a cartesian category. Let $\Delta=\pair{\id}{\id}: A\rightarrow A\& A$. Show that $\Delta$ and $\top_A$ define a commutative comonoid structure, i.e., verify the equations of exercise \ref{comonoid-laws-Linear}.
\end{exercise}
\begin{exercise} \label{comonoid-laws-Linear}
Let $d$ and $e$ be as just defined. Show that the following equations
are satisfied:
$$\begin{array}{lllllll}
\iota_l\comp(e\otimes\id)\comp d & = & \id &&
\iota_r\comp(\id\otimes e)\comp d & = & \id\\
\alpha\comp(\id\otimes d)\comp d & = & (d\otimes\id)\comp d &&
\gamma\comp d &= & d\; .
\end{array}$$
\end{exercise}

We are now in a position to sketch the interpretation of the sequents of linear logic
in a  Seely category $\cat{C}$. A proof of a sequent $\vdash 
A_1,\ldots,A_n$ is interpreted by a 
morphism $f:1\rightarrow(A_1\lpar\ldots\lpar A_n)$ (confusing the formulas 
with 
their interpretations as objects of $\cat{C}_l$), or (cf. exercise \ref{closed-dual-Linear}) as a morphism from $\Gamma^\bot$ to $\Delta$, for any splitting $A_1,\ldots,A_n=\Gamma\union\Delta$. We shall freely go from one of these representations to another.
The rules are interpreted as follows:

\Proofitem{(1)}  $\vdash 1$ is interpreted by $\id:1\rightarrow 1$.

\Proofitem{(\bot)} Obvious, since $\bot$ is the dual of $1$ (cf. 
exercise 
\ref{closed-dual-Linear}) and since $1\otimes 1\cong 1$.

\Proofitem{(\otimes)} From $f:\Gamma^{\bot}\rightarrow A$ and 
$g:\Delta^{\bot}\rightarrow B$, we form 
$f\otimes g:\Gamma^{\bot}\otimes\Delta^{\bot}\rightarrow A\otimes B$.

\Proofitem{(\lpar)} Obvious by associativity of $\lpar$.

\Proofitem{({\it Axiom})}  We simply take $id:A\rightarrow A$.

\Proofitem{({\it Cut})} Compose
$f:\Gamma^\bot \rightarrow A$ and $g:A\rightarrow\Delta$.

\Proofitem{(\top)} Interpreting $\vdash \top,\Gamma$ amounts to giving
an arrow from $\Gamma^{\bot}$ to $\top$. Since $\top$ is terminal, we take 
the unique such arrow.

\Proofitem{(\&)} The pairing of $f:\Gamma^{\bot}\rightarrow A$ and
$g:\Gamma^{\bot}\rightarrow B$ yields 
$\pair{f}{g}:\Gamma^{\bot}\rightarrow(A\& B)$.

\Proofitem{(\oplus)} Given $f:A^\bot\rightarrow\Gamma$, we build 
$f\comp\pi_{1}: A^{\bot}\& B^{\bot}\rightarrow\Gamma$, which we can 
consider as a morphism from $1$ to $(A\oplus B)\lpar\Gamma$.

\Proofitem{({\it Dereliction})} Given $f:A^\bot\rightarrow\Gamma$, we 
build
$f\comp\epsilon:\: !(A^{\bot})\rightarrow\Gamma$, where $\epsilon$ is the 
first 
natural transformation of the comonad.

\Proofitem{({\it Promotion})} Let 
$f:1\rightarrow(A\lpar ?B_1\lpar\ldots\lpar ?B_n)$, which we can consider as
an arrow from $!(B_1^{\bot}\&\ldots\& B_n^{\bot})$ to
$A$. Here we have made an essential use of the natural isomorphisms
required in the definition of a Seely category. Then we can 
consider
$\kleisli{f}:\: !(B_1^{\bot}\&\ldots\& B_n^{\bot})\rightarrow !A$
as a proof of $\vdash !A, ?B_1,\ldots, ?B_n$.

\Proofitem{({\it Weakening})} For this case and the following one,
we use the comonoid 
structure induced on $!A$.
Let $f:1\rightarrow\Gamma$. Then we can 
consider
that $f\comp e=\:!(A^{\bot})\rightarrow\Gamma$  is a proof of 
$\vdash ?A,\Gamma$.

\Proofitem{({\it Contraction})} This case is similar to the case 
$({\it Weakening})$, replacing $e$ by $d$.

\Proofitem{({\it Exchange})} By associativity and commutativity.

\medskip
So far, so good. But is the interpretation invariant under cut-elimination, which  is what we want of a model? The answer is:  almost. We shall treat the contraction reduction in detail. Referring to section \ref{full-ll} for notation, the proof $\Pi$ of $\vdash\Gamma_1,?B^\bot,?B^\bot$ is interpreted by, say, a morphism $h:!B\otimes\:!B\rightarrow C$, where $C$ interprets $\Gamma_1$, and the proof $\Pi'$ of $\vdash?\Gamma_2,B$ is interpreted by, say, a morphism $g:!A\rightarrow B$, where 
$A$ interprets $\Gamma_2^\bot$, assuming for the time being that $\Gamma_2$ is just one formula. Then the validation of the contraction reduction amounts to the following commutativity equation: $d_B\comp\: !g\comp \delta_A=((!g\comp\delta_A)\otimes(!g\comp\delta_A))\comp d_A$, which is a consequence of the following two commutativity equations:
$$\begin{array}{lll}
 d_{!A}\comp\delta_A=(\delta_A\otimes\delta_A)\comp d_A &&
(!g\otimes\:!g)\comp d_{!A}=d_B\comp !g\;.
\end{array}$$
These two equations are instances of the following more general statement:
$$\begin{array}{lll}
(P) & \qqs{f:!A\rightarrow !A'}{(\delta_{A'}\comp f=\:!f\comp\delta_A) & (\mbox{every free coalgebra morphism}\\
&\quad\implies
(d_{A'}\comp f=(f\otimes f)\comp d_A)} 
& \mbox{is also a comonoid morphism})
\end{array}$$
(A coalgebra over $A$ is a morphism $f:A\rightarrow !A$ and the free coalgebras are the
coalgebras $\delta_A$ over $!A$.)
Indeed, $\delta$ is a free coalgebra morphism, by the law $\delta_{!A}\comp\delta_A=\:!\delta_A\comp\delta_A$, and, for any $g:A\rightarrow A'$, $!g$ is a free coalgebra morphism by naturality of $\delta$.

Property $(P)$ cannot be derived from the sole axioms of Seely, as
was noted by Benton, Bierman, Hyland and de Paiva \cite{Bierman95} (for a survey and more references, see \cite{Mellies03}). Below, we show how to prove $(P)$ assuming only one new equation $(S^+)$ with respect to Seely's axiomatization. The reasoning could be carried out entirely in the category ${\bf C}$, but it really arises via a tour into the co-Kleisli adjunction of the comonad $!$ (cf. exercise \ref{comon-adj}), without which the equation $(S^+)$ would seem ad hoc.

The equation $(S^+)$ asserts the naturality of $n$, between the two functors $F\__1\otimes F\__2$
and $F(\__1\&\__2)$ from ${\bf C}_!\&{\bf C}_!$ to ${\bf C}$, that is,
for all $f\in{\bf C}_![A,A']$, $g\in{\bf C}_![B,B']$:
$$\begin{array}{ll}
(S^+) & F(f\& g)\comp n_{A,B}= n_{A',B'}\comp (Ff\otimes Fg)
\end{array}$$
or, in a form that only mentions ${\bf C}$:
$$\begin{array}{ll}
(S^+) & !<f\comp\:!\pi,g\comp\:!\pi'>\comp\:\delta_{A\& B}\comp n_{A,B}= n_{A',B'}\comp ((!f\comp\delta_A)\otimes (!g\comp\delta_B))\;.
\end{array}$$
Now we show that $(P)$ holds (we follow \cite{Mellies03}[section 3.5]). We first remark that $d$ can be reformulated as follows:
$$\begin{array}{lllll}
d & = & n^{-1}_{{\it UA},{\it UA}}\comp F\Delta_{{\it UA}} \\
& = & (\epsilon_{{\it FUA}}\otimes\epsilon_{{\it FUA}})\comp
(F\eta_{{\it UA}}\otimes F\eta_{{\it UA}}) \comp n^{-1}_{{\it UA},{\it UA}}\comp F\Delta_{{\it UA}} & (\mbox{adjunction law})\\
& = & (\epsilon_{{\it FUA}}\otimes\epsilon_{{\it FUA}})\comp n^{-1}_{{\it UFUA},{\it UFUA}}\comp
F(\eta_{{\it UA}}\&\eta_{{\it UA}})\comp F\Delta_{{\it UA}} & (\mbox{by }(S^+))\\
& = & (\epsilon_{{\it FUA}}\otimes\epsilon_{{\it FUA}})\comp n^{-1}_{{\it UFUA},{\it UFUA}}\comp
F\Delta_{{\it UFUA}}\comp F\eta_{{\it UA}} & (\Delta\mbox{ natural})
\end{array}$$

\noindent
As $F\eta_{{\it UA}}=\delta_A$, this gives us a handle to use the assumption of $(P)$ in the proof of its conclusion:
$$\begin{array}{ll}
d_{A'}\comp f\\
\; = (\epsilon_{{\it FUA}'}\otimes\epsilon_{{\it FUA}'})\comp n^{-1}_{{\it UFUA}',{\it UFUA}'}\comp
F\Delta_{{\it UFUA}'}\comp \delta_{A'}\comp f\\
\; =  (\epsilon_{{\it FUA}'}\otimes\epsilon_{{\it FUA}'})\comp n^{-1}_{{\it UFUA}',{\it UFUA}'}\comp
F\Delta_{{\it UFUA}'}\comp FUf\comp\delta_A & (\mbox{assumption})\\
\; =  (\epsilon_{{\it FUA}'}\otimes\epsilon_{{\it FUA}'})\comp n^{-1}_{{\it UFUA}',{\it UFUA}'}\comp
F(Uf\&Uf)\comp F\Delta_{{\it UFUA}}\comp\delta_A & (\Delta\mbox{ natural})\\
\; =  (\epsilon_{{\it FUA}'}\otimes\epsilon_{{\it FUA}'})\comp (FUf\otimes FUf)\comp n^{-1}_{{\it UFUA},{\it UFUA}}\comp F\Delta_{{\it UFUA}}\comp\delta_A & (\mbox{by }(S^+))\\
\; =  (f\otimes f)\comp(\epsilon_{{\it FUA}}\otimes\epsilon_{{\it FUA}})\comp n^{-1}_{{\it UFUA},{\it UFUA}}\comp F\Delta_{{\it UFUA}}\comp\delta_A & (\epsilon\mbox{ natural})\\
\; =( f\otimes f)\comp d_A
\end{array}$$

\smallskip
Now, we have to lift the restriction on the context $\Gamma_2$. What happens if $\Gamma_2$ consists of two or more formulas, say, $\Gamma_2=A_1^\bot,A_2^\bot$? Well, not a big deal, since 
$!A_1\otimes\:!A_2$ is isomorphic to $!(A_1\& A_2)$, so we can proceed as we did. Yes, but there
are two distinct contraction rules in
$$\seqdbl{\vdash\Gamma_1,?A_1^\bot,?A_2^\bot,?A_1^\bot,?A_2^\bot}{\vdash\Gamma_1,?A_1^\bot,?A_2^\bot}$$
and hence the interpretation of the right hand side is, up to associativity and commutavity,
$((!g\comp\delta_A)\otimes(!g\comp\delta_A))\comp (d_{A_1}\otimes d_{A_2})$ and {\em not}
$((!g\comp\delta_A)\otimes(!g\comp\delta_A))\comp d_{A_1\& A_2}$. We shall be done if we have 
$d_{A_1}\otimes d_{A_2}=d_{A_1\& A_2}$ up to associativity and commutativity. 

It remains to consider the case where $\Gamma_2$ is empty. Then one still proceeds as we did, setting $A=\top$, but at the price of introducing a useless $d:!\top\rightarrow\:!\top\otimes\:!\top$, which should be the identity up to the two canonical isomorphisms $\iota_l\comp(p^{-1}\otimes\:!\top)$ and  $\iota_r\comp(!\top\otimes p^{-1})$.
\begin{exercise} Formulate these equalities more precisely, and show that they are  consequences of the following equations (taken from \cite{Mellies03}[section 4])
$$\begin{array}{ll}
(S^+_\alpha) & n_{A\&B,C}\comp (n_{A,B}\otimes\:!C)\comp \alpha_{!A,!B,!C}=
\:!\alpha_{A,B,C}\comp n_{A,B\&C}\comp(!A\otimes n_{B,C})\\
(S^+_\gamma) & n_{B,A}\comp\gamma_{!A,!B}=\:!\gamma_{A,B}\comp n_{A,B}\\
(S^+_{\iota_l}) & (\iota_l)_{!A} = \:!(\iota_l)_A\comp n_{\top,A}\comp(p\otimes\:!A)\\
(S^+_{\iota_r}) & (\iota_r)_{!A} = \:!(\iota_r)_A\comp n_{A,\top}\comp(!A\otimes p)
\end{array}$$
which use the fact that a cartesian category is a fortiori monoidal.
\end{exercise}
\begin{exercise} With the help of the same equations, show that the weakening reduction and the ``box-box'' reduction (I.e. a cut between two promotions $\vdash ?\Gamma_1,?B^\bot,!C$ and
$\vdash?\Gamma_2,!B$) are valid. Hint: For the latter, prove first that
$\delta\comp Ff=\:!Ff\comp\delta$,  for any $f:!A\rightarrow B$.
\end{exercise}

We have completed the definition of a categorical model of linear logic, which we wrap in the following definition.
\begin{definition} 
A Seely$^+$ category is a Seely category which moreover satisifes $S^+$, $(S^+_\alpha)$,
$(S^+_{\iota_l})$, $(S^+_{\iota_r})$, and $(S^+_\gamma)$.
\end{definition}
In more synthetic terms, this definition says that the functor $F:\cat{C}_!\rightarrow \cat{C}$ is (strong) {\em monoidal}.

We end the section by sketching the proof that the category $\cat{Coh}_l$ of
coherence spaces and linear functions forms a Seely$^+$ category.
\begin{proposition}\label{coh-monclosed-Linear}
The inclusion functor from  $\cat{Coh}_l$  to the category \cat{Coh} of coherence spaces and stable functions has a left adjoint, and 
 $\cat{Coh}_l$ together with the comonad on $\cat{Coh}_l$ 
induced by the adjunction yields  a Seely category 
whose co-Kleisli category is equivalent to \cat{Coh}.
\end{proposition}
\Proof The adjunction is an immediate consequence of proposition \ref{birth-LL} and
exercise \ref{graph-linfun}.
For the symmetric monoidal structure, we just notice that 
at the level of events the 
canonical isomorphisms are given by:
$$\begin{array}{llll}
((e,e'),e'')  \leftrightarrow  (e,(e',e'')) &
(e,\ast)   \leftrightarrow  e &
(\ast,e)   \leftrightarrow  e & 
(e,e') \leftrightarrow (e',e) \; .
\end{array}$$
There is a natural bijection 
$\cat{Coh}_l[1,E]\cong D(E)$, since
 $(\ast,e_1)\coh(\ast,e_2)$ boils down to $e_1\coh e_2$. 
Hence we have:
$$\begin{array}{llll}
\cat{Coh}_l[1,(E\otimes E'^{\bot})^{\bot}] & \cong &
D(E\otimes E'^{\bot})^{\bot}) &\\
& = & D(E\limpl E') &\\
& \cong & \cat{Coh}_l[E,E'] & (\mbox{cf. exercise 
\ref{graph-linfun}})\; .
\end{array}$$
Then the closed structure follows from exercise 
\ref{dual-funct-Linear}.

\smallskip
To see that $\cat{Coh}_l$ is $\ast$-autonomous, we set 
$\bot=1^{\bot}\;(=1)$, 
and we observe the trace of
$\Lambda_l(\Lambda_l^{-1}(\id)\comp\gamma):A\rightarrow (A\limpl\bot)\limpl\bot$, which is
$\setc{(e,((e,\ast),\ast))}{e\in E}$.
It  has as inverse the function whose trace is
$\setc{(((e,\ast),\ast),e)}{e\in E}$.

That \cat{Coh} is equivalent to the co-Kleisli category follows from 
exercise
\ref{adj-com-Linear}.
We are left to verify the two natural isomorphisms. The first one 
holds by
exercise \ref{CCC-tensor-bang-Linear}. For the second one, notice 
that
$D(1)$ is a singleton.
\qed
\begin{exercise}
Show that the transformations $\epsilon$ and $\delta$ associated
with the comonad $!\:\comp\inc$ on $\cat{Coh}_l$ are the following
functions:
$$\begin{array}{lll}
\epsilon(X)=\setc{e}{\set{e}\in X} & &
\delta(X)=\setc{Y}{\Union Y\in X}\; ,
\end{array}$$
and check that the coherent model satisfies the other axioms of Seely$^+$ categories.
\end{exercise}

\end{document}